\documentclass[10pt,twocolumn,letterpaper]{article}

\usepackage{cvpr}              % To produce the CAMERA-READY version
\usepackage[dvipsnames]{xcolor}

\definecolor{cvprblue}{rgb}{0.21,0.49,0.74}
\usepackage[pagebackref,breaklinks,colorlinks,citecolor=cvprblue]{hyperref}
\usepackage[algoruled,vlined,linesnumbered]{algorithm2e}
\usepackage{graphicx}
\usepackage{array}
\newcolumntype{C}[1]{>{\centering}p{#1}}
\usepackage{amsmath}
\usepackage{amssymb}
\usepackage{booktabs}
\usepackage{enumerate}
\usepackage{arydshln}
\usepackage{bm}
\usepackage{cases}
\usepackage{color}
\usepackage{xcolor}
\usepackage{comment}
\usepackage{dsfont}
\usepackage{bbding}
\usepackage{tcolorbox}
\tcbuselibrary{theorems}
\usepackage{lipsum}
\usepackage{multirow}
\usepackage[pagebackref,breaklinks,colorlinks]{hyperref}

\def\paperID{6772} % *** Enter the Paper ID here
\def\confName{CVPR}
\def\confYear{2024}

\title{A Dynamic Kernel Prior Model for Unsupervised Blind Image Super-Resolution}

\author{Zhixiong Yang$^1$~~~~~~~~~~Jingyuan Xia$^{1,*}$~~~~~~~~~~Shengxi Li$^2$~~~~~~~~~~Xinghua Huang$^1$ \\
Shuanghui Zhang$^1$~~~~~~~~~~Zhen Liu$^1$~~~~~~~~~~Yaowen Fu$^1$~~~~~~~~~~Yongxiang Liu$^1$\\
\\
$^1$College of Electronic Engineering, National University of Defense Technology, Changsha, China\\
$^2$College of Electronic Engineering, Beihang University, Beijing, China\\
{\tt\small yzx21@nudt.edu.cn, j.xia10@nudt.edu.cn}
}

\definecolor{mycolor}{RGB}{206,255,255}    % {196,216,242} 
\definecolor{mycolor1}{RGB}{242,232,227}     %{242,232,227} 

\begin{document}
\maketitle
\footnotetext{Zhixiong Yang and Jingyuan~Xia contributed equally to this work (*Corresponding author: Jingyuan Xia).
This work is supported by National Natural Science Foundation of China, projects 61921001, 62131020, 62322121 and 62171448, and the NSFDYS of Hunan 2022J110067.
}
\begin{abstract}
Deep learning-based methods have achieved significant successes on solving the blind super-resolution (BSR) problem. However, most of them request supervised pre-training on labelled datasets.
This paper proposes an unsupervised kernel estimation model, named dynamic kernel prior (DKP), to realize an unsupervised and pre-training-free learning-based algorithm for solving the BSR problem. 
DKP can adaptively learn dynamic kernel priors to realize real-time kernel estimation, and thereby enables superior HR image restoration performances. 
This is achieved by a Markov chain Monte Carlo sampling process on random kernel distributions. 
The learned kernel prior is then assigned to optimize a blur kernel estimation network, which entails a network-based Langevin dynamic optimization strategy. 
These two techniques ensure the accuracy of the kernel estimation.
DKP can be easily used to replace the kernel estimation models in the existing methods, such as Double-DIP and FKP-DIP, or be added to the off-the-shelf image restoration model, such as diffusion model. 
In this paper, we incorporate our DKP model with DIP and diffusion model, referring to DIP-DKP and Diff-DKP, for validations. 
Extensive simulations on Gaussian and motion kernel scenarios demonstrate that the proposed DKP model can significantly improve the kernel estimation with comparable runtime and memory usage, leading to state-of-the-art BSR results. 
The code is available at \url{https://github.com/XYLGroup/DKP}.
% The proposed DKP consists of two main modules: random kernel sampling (RKS) module and prior kernel estimation (PKE) module.
% In the RKS module, we propose to model the kernel prior by using randomly sampled kernels to substitute conventional kernel priors.
% Based on the MCMC modeling, each randomly sampled kernel can be ensured to be rational.
% In the PKE module, the kernel is estimated by a shallow network.
% To ensure effective kernel estimation,
% we propose a network-based Langevin dynamics paradigm for the kernel estimation network update.
% This paradigm can incorporate kernel priors during the network update process while guarantee the data-consistent kernel.
% Finally, Our DKP model can achieve reliable kernel estimation by combining these two modules.
% For solving the BSR problem,
% we incorporate our DKP model into two image restoration models as two unsupervised BSR methods: DIP-DKP and Diff-DKP, which have no need of any pre-training/re-training nor network structures modification.
% Extensive simulations on both of Gaussian and motion kernel scenarios validate the superior performance and efficiency of our DIP-DKP and Diff-DKP. 
% An example code is given in the supplementary.
\end{abstract}    
\section{Introduction}
\label{sec:intro}

Deep learning provides a new avenue for solving the blind super-resolution (BSR) problem, which aims to reconstruct high-resolution (HR) images from the low-resolution (LR) observations with unknown blur kernels, and is known to be highly non-convex and ill-posed.
To alleviate the non-convexity and ill-posedness, 
most of learning-based BSR methods 
% \cite{dong2014learning, kim2016accurate, zhang2018learning, jin2018normalized, zhang2018image, liu2020residual, xu2020unified, zhang2020deep, hui2021learning, kim2021koalanet, wang2021unsupervised} 
incorporate image priors via supervised learning based on paired LR-HR samples. However, pre-defined labeled training datasets are expensive, time-consuming, and even not feasible in specific scenarios, such as for high speed targets (e.g., satellites, aircraft) and medical images (e.g., beating heart). 
% In these instances, this paper focuses on unsupervised learning-based solutions for the BSR problem.
Thus, unsupervised learning-based solutions are highly demanded for BSR problem.

The existing BSR methods can be roughly divided into model-based and learning-based strategies in terms of the priors adopted to provide performance guarantee.
Model-based approaches \cite{krishnan2009fast, kim2010single, perrone2015clearer, Yue2022blind} typically adopt hand-designed and explicit constraints as regularizations on image properties, or expert knowledge of the blur kernel.
Meanwhile, learning-based BSR methods \cite{zhang2018learning, zhang2018image, gu2019blind, liu2020residual, xu2020unified, zhang2020deep, huang2020unfolding, luo2021end, kim2021koalanet, luo2022deep} aim to train an end-to-end network with paired LR-HR image samples to leverage data priors for boosting performances. 
However, these methods highly demand the data and need to undergo throughing pre-training before applications, leading to limited generalization ability towards varying blur kernels. 
To alleviate this issue, quite a few methods \cite{song2022pseudoinverse, chung2022diffusion, zhao2023ddfm, guo2023shadowdiffusion, yi2023diff, zhu2023denoising} substitute the cumbersome training in advance by a well-trained diffusion model with significantly less fine-tuning samples in an off-the-shelf fashion. 
On the other side, a slice of works \cite{shocher2018zero, jin2018normalized, bell2019blind, ren2020neural, liang2021flow, Yue2022blind} propose to replace the HR image data priors by kernel priors, which are more substantial, economical and efficient to be trained. 
However, both of these advances are underlying the supervised learning scheme with necessity of training on labeled datasets, still hindering the flexibility and generalization ability towards the BSR tasks with different kernels and unknown HR ground truths. 

In this paper, we propose a dynamic kernel prior (DKP) generation model that can be plug-in with the majority of image restoration (IR) models, to solve BSR problem in an unsupervised way.
The proposed DKP model consists of two modules: random kernel sampling (RKS) module and prior kernel estimation (PKE) module. 
In the RKS module, a Markov Chain Monte Carlo (MCMC) sampling strategy on  kernel distributions iteratively generates random kernels as kernel priors, which are then assigned to the PKE module.
The PKE module is employed to estimate the blur kernel with respect to the kernel prior generated from the RKS module, the observed LR input and estimated HR image from the IR model. 
The estimated blur kernel is then assigned to an adopted IR model for the HR image restoration.
Along with the alternative solving processes, the MCMC process in RKS module converges to a desired kernel distribution with respect to the LR observation and the estimated HR image to guarantee a rational kernel prior. 
Meanwhile, a network-based Langevin dynamics (NLD) paradigm is proposed to optimize the kernel estimator in our PKE module with respect to the RKS output kernel prior and the data consistency based on the LR image reconstruction error.
The RKS module realizes an unsupervised kernel prior learning. 
The PKE module achieves promising kernel estimation via the NLD update scheme, which further alleviates the non-convexity and ill-posedness in the view of optimization strategy. 
In this way, the DKP model is capable of providing the plug-and-play kernel estimation without training in advance on paired LR-HR samples, and is flexible to be applied to the existing IR models for solving the BSR problem.
% For more effective sampling, a Markov Chain Monte Carlo (MCMC) modeling is adopted to enforce each sampled kernel to be rational.
% A set of random kernels are first sampled from the kernel prior distribution via Monte Carlo (MC) simulations.
% Then, the Markov Chain Monte Carlo (MCMC) modeling is adopted to enforce the randomly sampled kernels to be rational, which can be used as the rational kernel prior for the next kernel estimation.
% In the PKE module, we adopt a shallow network to estimate the kernel.
% A network-based Langevin dynamics update paradigm is proposed to update the network parameters, which consists of two update terms: the kernel prior term and data consistency term.
% The kernel prior term is computed based on the difference between the estimated kernel and the randomly sampled kernel from the RKS module, 
% while the data consistency term is computed based on the LR image reconstruction error.
% % difference between the given LR image and the estimated LR image that is blurred by the estimated kernel.
% Thus, this paradigm can embed kernel priors from RKS module during the network updating process while preserving the data-consistent kernel solution.
% Finally, by combining these two modules, the reliable kernel solution can be achieved by our DKP model.

Two applications are proposed to validate the feasibility and performance of our DKP model: deep image prior (DIP) \cite{ulyanov2018deep} and diffusion model \cite{ho2020denoising} adopted as the IR model, referring to DIP-DKP and Diff-DKP, respectively.
For the DIP-DKP, we simultaneously optimize the parameters of DIP and DKP models from scratch during the alternative solution process. 
For the Diff-DKP, the adopted diffusion model is off-the-shelf from \cite{ho2020denoising} and is applied as the fixed HR image restorer. The DKP model is optimized from scratch as well. 
Extensive simulation results show that the DIP-DKP achieves comparable performance than the existing methods, while the Diff-DKP achieves the state-of-the-art performance in both of Gaussian and motion kernel scenarios.
The main contributions are summarized as follows:

\begin{itemize}
\item 
The RKS module is proposed to generate a rational kernel prior from the MCMC sampling on random kernel distributions. This way, an unsupervised kernel prior learning is achieved to substitute the pre-training phase.
\item 
In PKE module, the NLD is proposed to optimize the kernel estimator, ensuring good convergence and concise estimation of the blur kernel from the perspective of optimization strategy.
\item 
The proposed DKP model enjoys the ease use on the popular IR models without the necessity of pre-training/re-training towards different scenarios. The two applications, i.e., DIP-DKP and Diff-DKP, validate the state-of-the-art performance and excellent flexibility of our DKP model.
\end{itemize}

\section{Related Work}\label{sec:Related Work}

% Existing BSR methods primarily focus on designing various image priors and kernel priors to address these two sub-problems in the original BSR problem: image restoration and kernel estimation, respectively.

% \subsection{Blind Super-resolution Methods}

% \noindent
% \textbf{Image Restoration.}
To alleviate the non-convexity and ill-posedness, early model-based approaches \cite{russell2003exploiting, glasner2009super, michaeli2013nonparametric, perrone2015clearer} typically construct image priors in explicit formulations, such as the total variation (TV) \cite{rudin1992nonlinear}, gradient profile \cite{sun2008image}, hyper-Laplacian \cite{krishnan2009fast} and sparsity \cite{kim2010single}. 
In contrast, learning-based methods \cite{dong2014learning, kim2016accurate, zhang2018learning, jin2018normalized, zhang2018image, gu2019blind, xu2020unified, huang2020unfolding, luo2021end, kim2021koalanet, wang2021unsupervised, luo2022deep} typically train an end-to-end network on labelled image samples to incorporate data priors. 
Wang \textit{et al}. \cite{wang2021unsupervised} proposed a CNN-based deep network with degradation feature representation module to learn image degradation feature from supervised training on paired LR-HR images.
Li \textit{et al}. \cite{li2023efficient} proposed a transformer network to learn multi-scale image feature via self-attention mechanisms. To reduce the high training costs of time and data, recent advances \cite{saharia2022image, song2022pseudoinverse, chung2022diffusion, wang2022zero, zhao2023ddfm, yi2023diff, zhu2023denoising} are proposed to solve BSR problem by an off-the-shelf diffusion model \cite{ho2020denoising}. 
Lin \textit{et al}. \cite{lin2023diffbir} proposed to partially fine-tune the parameters of diffusion model with significantly less labeled images. 
Wang \textit{et al}. \cite{wang2022zero} further formulated a diffusion-based BSR algorithm that iteratively solves super-resolution tasks with the given kernel without re-training. 
Different from the end-to-end models that are trained on paired image samples, recent methods tend to resolve BSR problem via pre-training on kernel datasets \cite{liang2021flow} or pre-defined kernel priors \cite{Yue2022blind}. An alternative framework between the kernel estimation and image restoration is typically adopted in these methods \cite{ulyanov2018deep, shocher2018zero, gandelsman2019double, bell2019blind, gu2019blind, zhou2023learning}, such as double-deep image prior (Double-DIP) \cite{ren2020neural}.
On the basis of this framework, Liang \textit{et al}. \cite{liang2021flow} established a flow-based kernel prior (FKP) network that is pre-trained on labeled kernels to enroll kernel priors while the HR image is estimated by DIP network in an online fashion. 
Yue \textit{et al}. \cite{Yue2022blind} proposed a hand-crafted kernel prior model to improve the robustness towards the Gaussian kernel scenario. 
Despite the fact that these methods approximately bring down the data requirements and training costs, the necessity of training in advances or hand-crafted design still limits the flexibility and generalization ability towards the varying kernel scenarios (Gaussian and motion) without ground truths. 

\section{Dynamic Kernel Prior (DKP)}\label{sec:DKP}
% \subsection{Dynamic Kernel Prior}

\noindent
\textbf{Problem Formulation.}
The degradation model of BSR problem is commonly expressed as follows,
\begin{equation}
\bm{y}=(\bm{x}\otimes \bm{k})\downarrow _s+\bm{n}, \label{eq:degradation model}
\end{equation} 
where $\bm{y}$ denotes the LR image, $\bm{x}$ denotes the HR image, $\otimes$ indicates the convolution operation, $\downarrow _s$ denotes the down-sampling operation with scale factor $s$, and $\bm{k}$ denotes the blur kernel.
% Notably, different types of $\bm{k}$ correspond to different image super-resolution tasks, such as the Gaussian blur and motion blur.
The BSR problem \eqref{eq:degradation model} can be formulated as a maximum a posteriori (MAP) problem:
\begin{equation} \label{eq:BSR}
\;\underset{\bm{x},\bm{k}}{\max} \; p_{}(\bm{y}|\bm{x},\bm{k}) {p}_{}(\bm{x}) {p}_{}(\bm{k}),
\end{equation}
where $p_{}(\bm{y}|\bm{x},\bm{k})$ denotes the likelihood of the observed LR image $\bm{y}$, ${p}(\bm{x})$ and ${p}(\bm{k})$ are the HR image and kernel priors, respectively.
Image priors \cite{ulyanov2018deep, song2019generative, ho2020denoising, zhang2021plug, dong2022unpaired} have been well-designed and fully-studied in the past decade.
In contrast, researches on kernel priors ${p}(\bm{k})$ are in the ascendant, as kernel samples are less expensive to obtain and the training phase is more efficient \cite{efrat2013accurate, zhang2018learning, gu2019blind, liang2021flow, Yue2022blind}.

In this paper, we propose the DKP model, which comprises two modules: RKS and PKE. The RKS module is employed to generate rational kernel priors, which are assigned to the PKE module to support the estimation of blur kernel. Let $t=1,2,\cdots,T$ denote the alternative iterations among these two modules and the adopted IR model, $\bm{k}^t$ and $\bm{x}^t$ denote the estimated blur kernel and HR image at the $t^{\textit{th}}$ iteration, respectively. The details of DKP model is given below.

% Further details regarding these modules will be elaborated in the subsequent part of this section.

\begin{figure}[t]
\vspace{-0.2cm}
  \centering
  \includegraphics[width=0.82\linewidth]{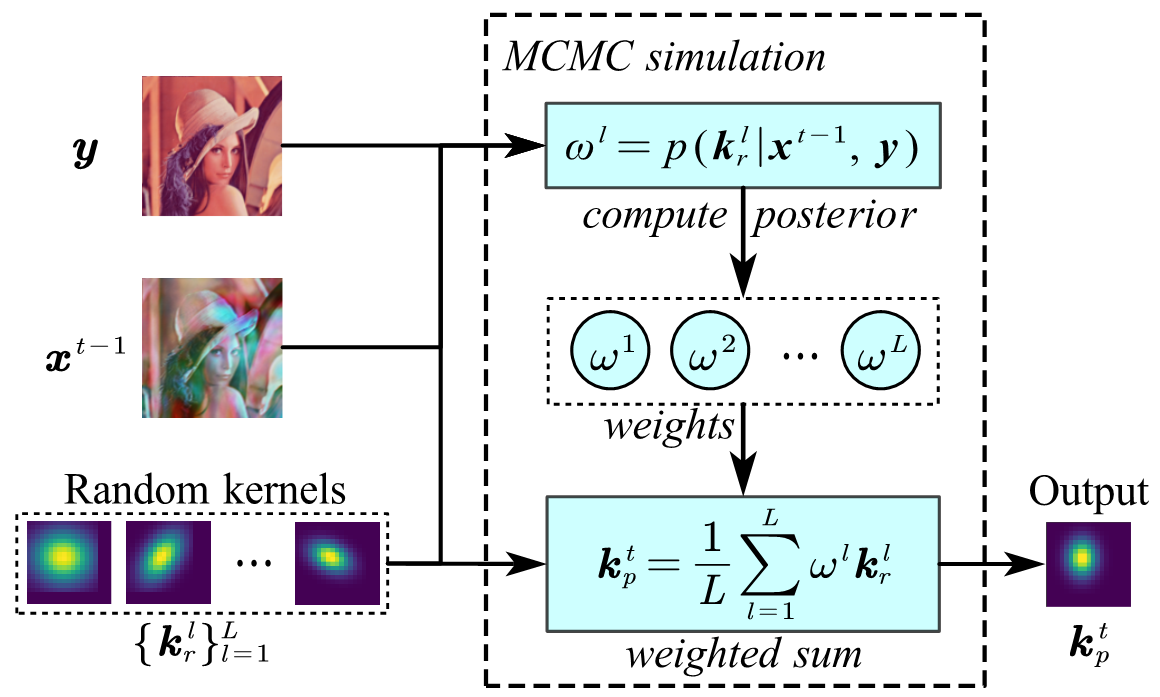}\\
  \vspace{-0.2cm}
  \caption{The overview of the RKS module. The MCMC simulation can generate the random kernel ${\bm{k}}^t_p$ from random kernel distributions $\{{\bm{k}}_r^{l}\}_{l=1}^L$ as the kernel prior with respect to the current model parameters $\bm{x}^{t-1}, \bm{y}$.
  }\label{fig: RKS}
  \vspace{-0.2cm}
\end{figure}

\noindent
\textbf{RKS module.}
At the $t^{\textit{th}}$  iteration, the RKS module plays the key role of generating a rational kernel prior ${\bm{k}}^t_p$ from the MCMC simulation. The overview diagram is shown in Fig. \ref{fig: RKS}. 
% learn a dynamic disturbation $\zeta_{\bm{\phi}_{\bm{k}},l}$ achieved by using Monte Carlo (MC) simulations to obtain samples that follow the prior distribution $\hat{p}_{K}(\bm{k})$.
% As shown in Fig. \ref{fig: DKP}. The estimated kernel $\bm{k}$ obeys the kernel distribution $p_{\bm{k}}(\bm{k})$.
% We sample random kernels $\hat{\bm{k}}_r$ from the prior distribution $\hat{p}_{K}(\bm{k})$ via MC simulations.
% Then, by calculating the loss between $\bm{k}$ and $\hat{\bm{k}}_r$ to update $\text{G}_{\bm{k}}$, the disturbation $\zeta_{\bm{\phi}_{\bm{k}}}$ of the random kernels $\hat{\bm{k}}_r$ can be incorporated into the network update process, while the kernel prior $\hat{p}_{K}$ is emerged.
Let $p({\bm{k}}_r|{\bm{\Sigma}}_{r})$ denotes that the random kernel ${\bm{k}}_r$ is conditioned by the latent variable ${\bm{\Sigma}}_{r}$, in which  $p({\bm{\Sigma}}_{r})$ determines the category of blur kernel.
Then the distribution of the kernel prior ${\bm{k}}^t_p$ can be formulated as
\begin{equation}\label{eq:random kernel expectation}
{p}({\bm{k}}^t_p) = \int_{{\bm{\Sigma}}_{r}}p({\bm{k}}_r|{\bm{\Sigma}}_{r})p({\bm{\Sigma}}_{r})d{\bm{\Sigma}}_{r}.
\end{equation}
Here, ${\bm{\Sigma}}_{r}$ is the parameter of kernel (e.g., the variance of Gaussian kernel or the length of motion kernel).
It is not easy to sample all the possible ${\bm{\Sigma}}_{r}$, and therefore, we convert \eqref{eq:random kernel expectation} into the Monte Carlo simulation in the following form:
% The expectation of random sample $p(\bm{k}_r)$ can be obtained by calculating the integration of $\bm{\Sigma}$ in the following:
\begin{equation}\label{eq:random kernel MC}
{p}({\bm{k}}^t_p) \approx \sum^L_{l=1} p({\bm{k}}_r^{l}|{\bm{\Sigma}}_r^{l}){\bm{\Sigma}}_r^{l},
\end{equation}
where $l$ denotes the index of the Monte Carlo sampling, ${\bm{\Sigma}}_r^{l}$ denotes the $l^{th}$ sampled latent variable, ${\bm{k}}_r^{l}$ is the $l^{th}$ sampled kernel, conditioned on the ${\bm{\Sigma}}_r^{l}$.
To ensure the rationality of randomly generated kernels towards the BSR problem, as well as the optimization during the iterations, the MCMC simulation is proposed as follows,
\begin{align}\label{eq:task-specific kernel MC}
{p}({\bm{k}}^t_p|{\bm{x}}^{t-1},\bm{y}) &\approx \sum^L_{l=1} p({\bm{k}}_r^{l}|{\bm{x}}^{t-1},\bm{y}) p({\bm{k}}_r^{l}| {\bm{\Sigma}}_r^{l}){\bm{\Sigma}}_r^{l},
\end{align}
where $p({\bm{k}}_r^{l}|{\bm{x}}^{t-1},\bm{y})$ denotes the kernel weight $\omega^l$, that is conditioned on the observed LR image $\bm{y}$ and the estimated HR image $\bm{x}^{t-1}$ with respect to the  MCMC loss $\mathcal{L}_{\textit{MCMC}}$ in the following form
\begin{equation}\label{eq:MCMC weight}
\omega^l = p({\bm{k}}_r^{l}|{\bm{x}}^{t-1},\bm{y}) \propto \frac{1}{\mathcal{L}^l_{\textit{MCMC}}},
\end{equation}
where 
\begin{equation}\label{eq:MCMC weight loss}
\mathcal{L}^l_{\textit{MCMC}} = \|\bm{y}-({\bm{x}}^{t-1}\otimes{\bm{k}}_r^{l})\downarrow _s\|_F^2 + \delta,
% + \|\text{G}_{\bm{k}}(\bm{\phi}_{\bm{k},l}) - \hat{\bm{k}}_{r,l})\|_F^2
\end{equation}
% ${\bm{k}}_r^{l}$ is the kernel determined by ${\bm{\Sigma}}_r^{l}$, 
$\delta$ is the noise to prevent $\mathcal{L}^l_{\textit{MCMC}}=0$. 
% Thus, the vanilla MC sampling of ${p}(\hat{\bm{k}}^t_r)$ in Eq. \eqref{eq:random kernel MC} is replaced by the MCMC sampling of ${p}({\bm{k}}^t_r|{\bm{x}}^{t-1},\bm{y})$ in Eq. \eqref{eq:task-specific kernel MC}.
In this way, ${\bm{k}}^t_p$ can be formulated as
\begin{equation}\label{eq:reweighted random kernel}
{\bm{k}}^t_p = \frac{1}{L} \sum_{l=1}^L \omega^l {\bm{k}}_r^{l}.
\end{equation}
The obtained ${\bm{k}}^t_p$ is then assigned to the PKE module as a rational kernel prior, which will be introduced next.

We note that the obtained kernel prior $\bm{k}^{t}_p$ is an expectation of $L$ times sampling according to \eqref{eq:random kernel MC}. 
The number of the sampling times $L$ plays the role of annealing/tempering in MCMC simulations as a hyper-parameter. Details of the tuning on  $L$ will be given in Section \ref{sec:setup}.
% In practice, we tune the value of $L$ instead of estimating it, and the tuning comparison is given in Section \ref{sec:setup}. 
% Furthermore, the sampled kernel $\bm{k}_r^{t}$ in the RKS module is acquired in real-time rather than pre-training. 

% along with the alternating between kernel estimation and image restoration,
% the sampled kernel $\hat{\bm{k}}^t_r$ of the MCMC simulation will converge to a desired kernel distribution with respect to the $\bm{y}$ and $\bm{x}^{t-1}$.

% the sampled kernel in the RKS module is acquired in real-time rather than pre-training. 
% The sampled kernel is then utilized as the kernel prior and integrated into the next PKE module for improved kernel estimation, which will be detailed next.

\noindent
\textbf{PKE module.}
In our DKP model, the PKE module is employed to estimate the blur kernel by a lightweight network $\text{G}_{\bm{k}}$ with parameters $\bm{\phi}_{\bm{k}}$ as follows 
\begin{equation}\label{eq:generate k}
{\bm{k}^{t}} = \text{G}_{\bm{k}}(\bm{\phi}^{t}_{\bm{k}}).
\end{equation}
The network $\text{G}_{\bm{k}}$ takes a fixed noise that is randomly initialized as input, and we neglect it for demonstration convenience as $\bm{\phi}^{t}_{\bm{k}}$ are the main variables. 

\begin{figure}[t]
\vspace{-0.2cm}
  \centering
  \includegraphics[width=1\linewidth]{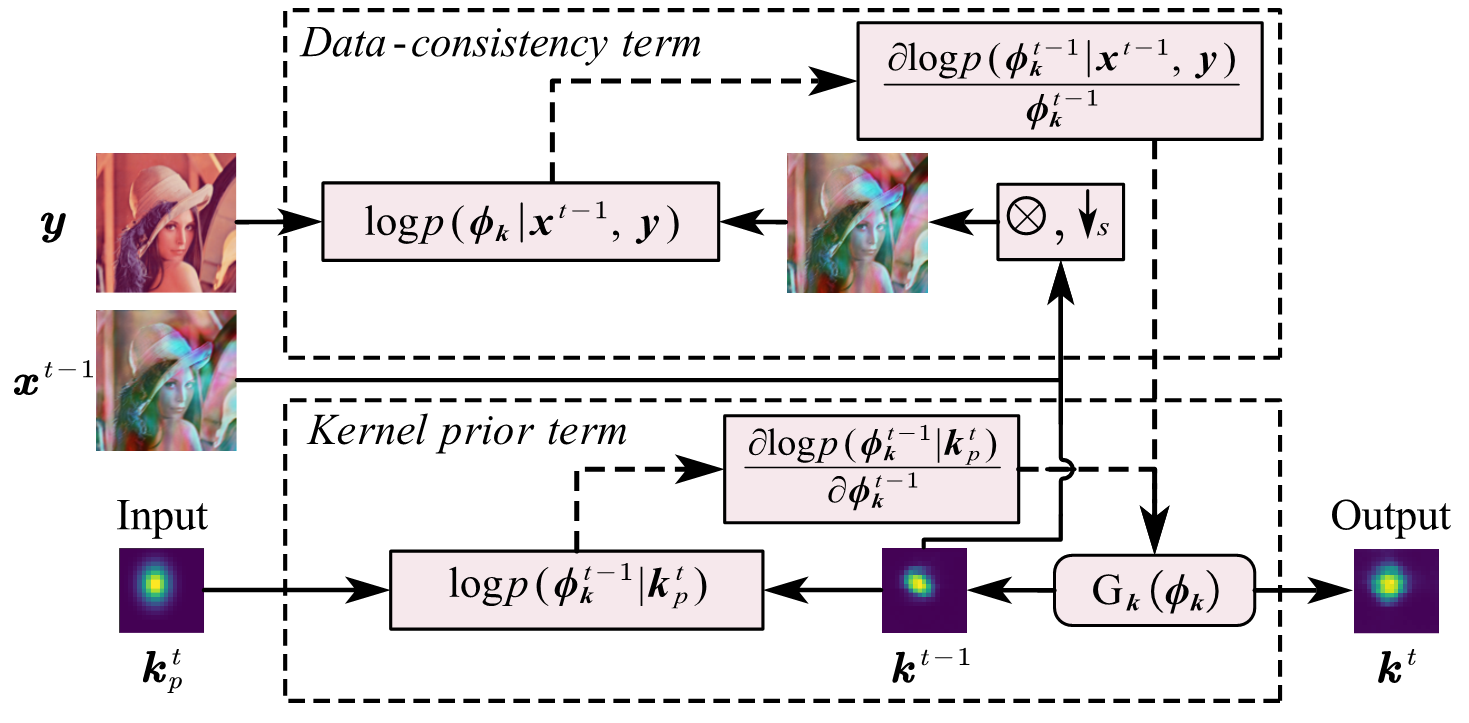}\\
  \vspace{-0.2cm}
  \caption{The overview of the PKE module. The blur kernel $\bm{k}^t$ is estimated by the network $G_{\bm{k}}$, whose parameters $\bm{\phi}_{\bm{k}}$ are updated by the kernel prior term from RKS module and data-consistency term, based on the LR image reconstruction error.
  }\label{fig: PKE}
  \vspace{-0.2cm}
\end{figure}

This kernel estimator $\text{G}_{\bm{k}}$ is optimized in the NLD paradigm with respect to the data-consistency term and kernel prior term, as shown in Fig. \ref{fig: PKE}. 
The data-consistency term is computed by the LR image reconstruction error, which is given by 
\begin{equation}\label{eq:data posterior term}
\log p_{\textit{}}( \bm{\phi}^{t-1}_{\bm{k}}|{\bm{x}}^{t-1},\bm{y}) = - \|\bm{y}-({\bm{x}}^{t-1}\otimes\text{G}_{\bm{k}}(\bm{\phi}^{t-1}_{\bm{k}}))\downarrow _s\|_F^2.
\end{equation}
The kernel prior term is computed based on the difference between the network-estimated kernel $\text{G}_{\bm{k}}(\bm{\phi}^{t-1}_{\bm{k}})$ and the random-sampled kernel ${\bm{k}}^t_p$ from the RKS module as follows,
\begin{equation}\label{eq:random kernel update term}
\log p_{\textit{}}(\bm{\phi}^{t-1}_{\bm{k}}|{\bm{k}}^t_p) = - \|\text{G}_{\bm{k}}(\bm{\phi}^{t-1}_{\bm{k}}) - {\bm{k}}^t_p\|_F^2.
\end{equation}
By combining \eqref{eq:data posterior term} and \eqref{eq:random kernel update term}, the network parameters $\bm{\phi}^{t-1}_{\bm{k}}$ can be updated as follows,
\begin{align}\label{eq:network update via LD}
{\bm{\phi}}^{t}_{\bm{k}} = \bm{\phi}^{t-1}_{\bm{k}} + \frac{\delta^2}{2}  & \frac{\partial \log p_{\textit{}}( \bm{\phi}^{t-1}_{\bm{k}}|{\bm{x}}^{t-1},\bm{y})}{\partial \bm{\phi}^{t-1}_{\bm{k}}}\notag\\
& \;\;\;\;\;\;\;+ \delta \frac{\partial \log p_{\textit{}}(\bm{\phi}^{t-1}_{\bm{k}}|{\bm{k}}^t_p)}{\partial \bm{\phi}^{t-1}_{\bm{k}}},
\end{align}
where the second term is the data-consistency update,
the third term is the additional update based on the random kernel ${\bm{k}}^t_p$.

It has been proved to be effective that the random noise-based disturbation can prevent being trapped into bad local modes for the variable update in Langevin dynamics \cite{bakry2006diffusions, neal2011mcmc, welling2011bayesian, Yue2022blind}. 
More details of Langevin dynamics refer to the supplementary material.
At this stage, the random kernel sample from the RKS module can be regarded as the random ``noise" for the $\bm{\phi}^{t-1}_{\bm{k}}$ update. 
Eq. \eqref{eq:network update via LD} can be reformulated as follows,
\begin{align}\label{eq:network update via NLD}
{\bm{\phi}}^{t}_{\bm{k}} = \bm{\phi}^{t-1}_{\bm{k}} + \frac{\delta^2}{2} & \frac{\partial \log p_{\textit{}}( \bm{\phi}^{t-1}_{\bm{k}}|{\bm{x}}^{t-1},\bm{y})}{\partial \bm{\phi}^{t-1}_{\bm{k}}} + \zeta^{t-1}_{\bm{\phi}_{\bm{k}}},
\end{align}
where $\zeta^{t-1}_{\bm{\phi}_{\bm{k}}} = \frac{\partial \log p_{\textit{}}(\bm{\phi}^{t-1}_{\bm{k}}|{\bm{k}}^t_p)}{\partial \bm{\phi}^{t-1}_{\bm{k}}}$ denotes the parameters correlated Langevin dynamics disturbation.
% Finally,

% Finally, by substituting the Eq. \eqref{eq:random kernel update term} and Eq. \eqref{eq:data posterior term} into Eq. \eqref{eq:network update via NLD}, we can obtain the explicit NLD formulation to compute the updated network parameters ${\bm{\phi}}^{t}_{\bm{k}}$.
% Then, the final estimated kernel $\bm{k}^{t} = \text{G}_{\bm{k}}({\bm{\phi}}^{t}_{\bm{k}})$ can be obtained by the updated network parameters ${\bm{\phi}}^{t}_{\bm{k}}$ via Eq. \eqref{eq:generate k}.

% However, it is clear that a Gaussian-distributed will not contribute to the $\bm{\phi}_{\bm{k}}$ optimization, as the parameters posterior $p_{}( \bm{\phi}_{\bm{k}}^{t-1}|\bm{x}^{t-1}, \bm{y})$ is implicit and entails data-based training. 

The pipeline of our DKP at the $t^{th}$ iteration is given in Algorithm \ref{alg: DKP}.
The whole DKP model is implemented in a plug-and-play style, in which training in advance is not required. 
Besides, the random kernels from the RKS module are self-adaptively sampled through the MCMC simulation, without the need of labeled training data. We should also note that the DKP model only brings neglectable run-time and memory cost in applications, as the adopted network $\text{G}_{\bm{k}}$ is typically lightweight. This leads to high flexibility and low computational complexity.
These three merits promise our DKP the convenience of being applied to the existing image restoration approaches, including the untrained DIP model and off-the-shelf pre-trained diffusion model, which will be detailed in the next section.

\begin{algorithm}[t]
    \SetAlgoLined
    \footnotesize
    \textbf{Given:} ${\bm{x}}^{t-1},\bm{y}$ and $\bm{\phi}^{t-1}_{\bm{k}}$.

    \% Random Kernel Sampling (RKS) Module
    
    Sample random kernels $\{{\bm{k}}_r^{l}\}_{l=1}^L$ via MC.
    
    \For{l $\gets$ 0, 1, $\ldots$, L}{

    $\omega^l = \frac{1}{\mathcal{L}^l_{\textit{MCMC}}}$, $\mathcal{L}^l_{\textit{MCMC}} = \|\bm{y}-({\bm{x}}^{t-1}\otimes{\bm{k}}_r^{l})\downarrow _s\|_F^2 + \delta$ %(Eq. \eqref{eq:MCMC weight loss})

     %(Eq. \eqref{eq:MCMC weight})

    }

    ${\bm{k}}^t_p = \frac{1}{L} \sum_{l=1}^L \omega^l {\bm{k}}_r^{l}$ %(Eq. \eqref{eq:reweighted random kernel})

    % $\log p_{\textit{}}(\bm{\phi}^{t-1}_{\bm{k}}|\hat{\bm{k}}^t_r) = \|\text{G}_{\bm{k}}(\bm{\phi}^{t-1}_{\bm{k}}) - \hat{\bm{k}}^t_{r})\|_F^2$ %(Eq. \ref{eq:random kernel prior term})

    % $\log p_{\textit{}}( \bm{\phi}^{t-1}_{\bm{k}}|{\bm{x}}^{t-1},\bm{y}) = \|\bm{y}-({\bm{x}}^{t-1}\otimes\text{G}_{\bm{k}}(\bm{\phi}^{t-1}_{\bm{k}}))\downarrow _s\|_F^2$ % (Eq. \eqref{eq:data posterior term})

    \% Prior Kernel Estimation (PKE) Module

    $ {\bm{\phi}}^{t}_{\bm{k}} = \bm{\phi}^{t-1}_{\bm{k}} + \frac{\delta^2}{2}  \frac{\partial \log p_{\textit{}}( \bm{\phi}^{t-1}_{\bm{k}}|{\bm{x}}^{t-1},\bm{y})}{\partial \bm{\phi}^{t-1}_{\bm{k}}} + \delta \frac{\partial \log p_{\textit{}}(\bm{\phi}^{t-1}_{\bm{k}}|{\bm{k}}^t_p)}{\partial \bm{\phi}^{t-1}_{\bm{k}}}$
    
    \textbf{Output:} ${\bm{k}}^t = \text{G}_{\bm{k}}({\bm{\phi}}^t_{\bm{k}})$.
\caption{\label{alg: DKP} The proposed DKP model.}
\vspace{-0.02in}
\end{algorithm}

\section{DKP-based BSR Methods}\label{sec:SR Module}
% Existing off-the-shelf image restoration modules can be roughly divided into two categories: DIP-based \cite{ulyanov2018deep, gandelsman2019double, ren2020neural, liang2021flow, Yue2022blind} and diffusion-based \cite{saharia2022image, wang2022zero, zhao2023ddfm, guo2023shadowdiffusion, yi2023diff, lin2023diffbir}. 
% \subsection{Super-resolution Module}

\begin{figure}[t]
\vspace{-0.2cm}
  \centering
  \includegraphics[width=1\linewidth]{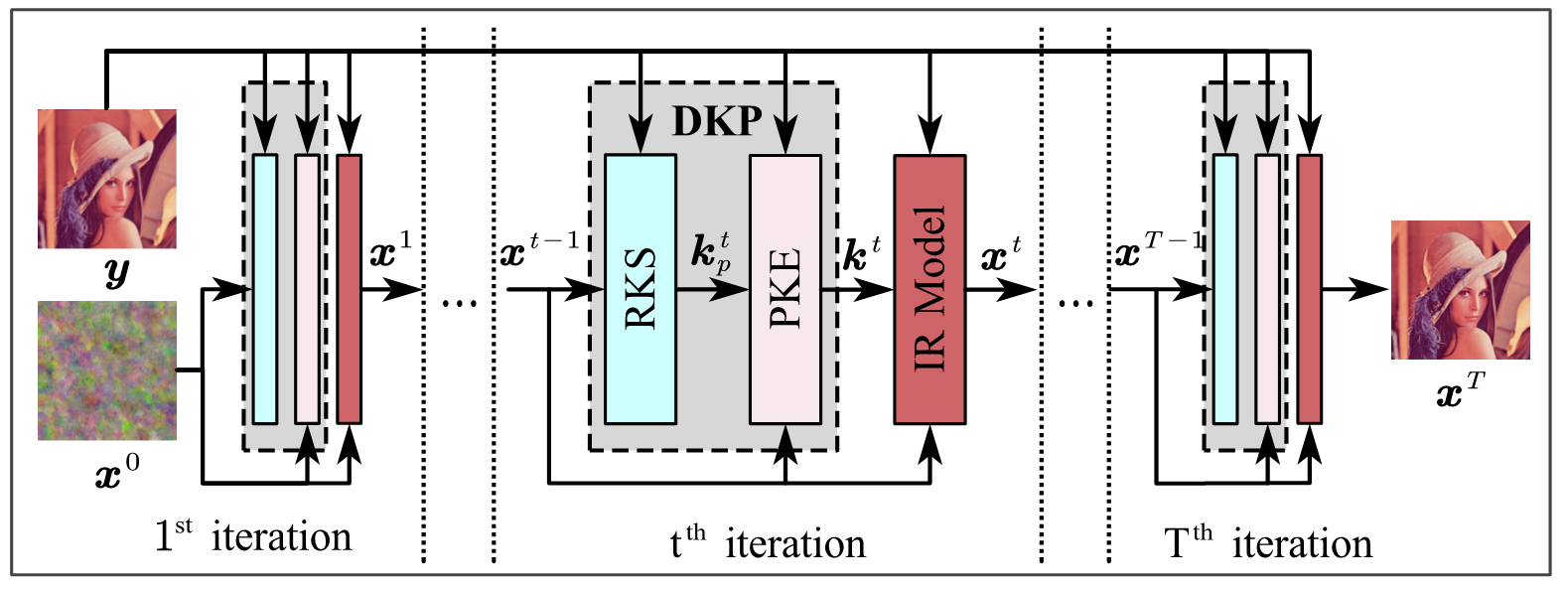}\\
  % \vspace{-0.16in}
  \caption{The overview of our DKP-based BSR method. 
  }\label{fig: DKP based BSR}
  \vspace{-0.2cm}
\end{figure}

\subsection{Pipeline}
The overview of the proposed DKP-based BSR method is illustrated in Fig. \ref{fig: DKP based BSR}.
The DKP model (gray box), including RKS module (blue box), PKE module (lilac box), and IR model (red box) alternatively optimize the blur kernel and refine the HR image, respectively.
% Specifically, at the $t^{th}$ iteration, the RKS module first generate a randomly sampled kernel $\hat{\bm{k}}^t_r$, which is assigned to the PKE module.
% The estimated kernel $\bm{k}^t$ is then output by the PKE module with respect to the $\hat{\bm{k}}^t_r$, $\bm{x}^{t-1}$ and $\bm{y}$.
For each iteration, the estimated HR image $\bm{x}^{t-1}$ and LR image $\bm{y}$ are first fed to RKS module ${f}_{\text{RKS}}$ to generate kernel prior
% \begin{equation}\label{eq:DKP module}
% {\bm{k}}^t = {f}_{\text{DKP}}\left({\bm{x}}^{t-1},\bm{y}\right),
% \end{equation}
\begin{equation}\label{eq:RKS module}
{\bm{k}}_p^t = {f}_{\text{RKS}}\left({\bm{x}}^{t-1},\bm{y}\right),
\end{equation}
where ${\bm{x}}^{t-1}$ denotes the estimated HR image from the last IR model output.
Then, the kernel prior ${\bm{k}}_p^t$ will be assigned to the PKE module ${f}_{\text{PKE}}$, which estimates kernel as follows,
\begin{equation}\label{eq:PKE module}
{\bm{k}}^t = {f}_{\text{PKE}}\left({\bm{x}}^{t-1},\bm{y},{\bm{k}}_p^t\right),
\end{equation}
where ${\bm{k}}^t$ is the estimated kernel at the $t^{th}$ kernel estimation iteration, which will be assigned to the IR model.
The $t^{th}$ HR image ${\bm{x}}^t$ can be estimated by the IR model as follows
\begin{equation}\label{eq:IR model}
{\bm{x}}^t = {f}_{\text{IR}}\left({\bm{x}}^{t-1},\bm{y},{\bm{k}}^t\right),
\end{equation}
where ${f}_{\text{IR}}$ denotes the adopted IR model.
In this paper, two representative IR models, DIP \cite{ulyanov2018deep} and diffusion model \cite{ho2020denoising} , are applied to evaluate the DKP-based BSR solutions, referring to DIP-DKP and Diff-DKP, which are introduced in the sequel.

\subsection{The proposed DIP-DKP}

\noindent
\textbf{DIP-based Image Restoration.}
DIP \cite{ulyanov2018deep} is designed for capturing low-level image statistics, and estimates HR image $\bm{x}=\text{G}_{\bm{x}}(\bm{z}_{\bm{x}},{\bm{\phi}}_{\bm{x}})$ from a fixed random noise input $\bm{z}_{\bm{x}}$ (we omit $\bm{z}_{\bm{x}}$ in the rest of this paper for demonstration convenience). A typical formulation of DIP-based BSR methods \cite{ren2020neural, liang2021flow} is given as follows 
% \begin{numcases}
% \;{\bm{\phi}}^{t}_{\textit{DIP}} = {\bm{\phi}}^{t-1}_{\textit{DIP}} + \gamma^{t-1}_{\textit{DIP}} \cdot \frac{\partial \log p_{\textit{}}( \bm{\phi}^{t-1}_{\textit{DIP}}|\bm{y})}{\partial \bm{\phi}^{t-1}_{\textit{DIP}}},\label{eq:DIP loss}\\
% {\bm{x}}^{t} = \text{G}_{\textit{DIP}}({\bm{\phi}}^{t}_{\textit{DIP}}),\label{eq:DIP-x}
% \end{numcases}
\begin{numcases}
\;{\bm{\phi}}^*_{\bm{x}}, {\bm{\phi}}^*_{\bm{k}} = \underset{\bm{\phi}_{\bm{x}}, {\bm{\phi}}_{\bm{k}}}{\arg \min}\;\|\bm{y}- \notag\\
\;\;\;\;\;\;\;\;\;\;\;\;\;\;\;\;\;\;\;\;\;\;\;\;(\text{G}_{\bm{x}}(\bm{\phi}_{\bm{x}})\otimes \text{G}_{\bm{k}}({\bm{\phi}}_{\bm{k}}))\downarrow _s\|_F^2,\label{eq:DIP loss} \\
{\bm{x}}^* = \text{G}_{\bm{x}}({\bm{\phi}}^*_{\bm{x}}), {\bm{k}}^* = \text{G}_{\bm{k}}({\bm{\phi}}^*_{\bm{k}}). \label{eq:DIP-x}
\end{numcases}

% \begin{equation}\label{eq:DIP network update}
% \log p_{\textit{}}( \bm{\phi}^{t-1}_{\textit{DIP}}|\bm{y}) = \|\bm{y}-\mathcal{D}\left(\text{G}_{\textit{DIP}}(\bm{\phi}^{t-1}_{\textit{DIP}})\right)\|_F^2,
% \end{equation}
% $\mathcal{D}(\cdot)$ denotes the known image degradations, such as bicubic downsampling, 
% $\text{G}_{\textit{DIP}}$ denotes the DIP network, $\bm{\phi}_{\textit{DIP}}$ denotes the network parameters, the network input is omitted, which is usually set to be fixed during the update process.

Double-DIP \cite{ren2020neural} and FKP-DIP \cite{liang2021flow} have exploited the effectiveness towards the BSR problem. However, the kernel prior of $\text{G}_{\bm{k}}({\bm{\phi}}^*_{\bm{k}})$ either adopts the untrained network with limited performances on the kernel estimation \cite{ren2020neural}, or pre-trained kernel network, referring to FKP \cite{liang2021flow}, that requests supervised training in advance. 
As shall be shown in experiments, pre-trained networks do not perform well to generate reasonable kernel estimations when the kernel categories vary.  
% By repeating the above process T times, the estimated image can be obtained through the updated DIP network via Eq. \eqref{eq:DIP-x}.

\begin{algorithm}[t]
    \SetAlgoLined
    \footnotesize
    \textbf{Given:} $\bm{y}$, $\bm{\phi}^0_{\bm{x}}$, $\bm{\phi}^0_{\textit{DKP}}$, ${\bm{x}}^0 = \text{G}_{\bm{x}}({\bm{\phi}}^0_{\bm{x}})$.
    
    % \textbf{Initialize:} 
    
    \For{t $\gets$ 0, 1, $\ldots$, T-1}{

    \begin{minipage}[t][0.4cm][t]{0.42\textwidth} % 设置minipage的高度为4cm
        \colorbox{mycolor}{%
            \begin{minipage}{\dimexpr\textwidth-2\fboxsep} % 设置minipage的宽度
                \% DKP-based kernel estimation stage
            \end{minipage}%
        }
    \end{minipage}
    
    \vspace{-0.08em}
    
    % \begin{minipage}[t][0.4cm][t]{0.42\textwidth} % 设置minipage的高度为4cm
    %     \colorbox{mycolor}{%
    %         \begin{minipage}{\dimexpr\textwidth-2\fboxsep} % 设置minipage的宽度
    %             \State  Sample random kernels $\{{\bm{k}}_r^{l}\}_{l=1}^L$ via MC.
    %         \end{minipage}%
    %     }
    % \end{minipage}
    
    % \begin{minipage}[t][0.5cm][t]{0.42\textwidth} % 设置minipage的高度为4cm
    %     \colorbox{mycolor}{%
    %         \begin{minipage}{\dimexpr\textwidth-2\fboxsep} % 设置minipage的宽度
    %             ${\bm{k}}^t = {f}_{\text{DKP}}\left({\bm{x}}^{t-1},\bm{y}\right)$ 
    %             % $\;$ (Eq. \eqref{eq:DKP model in DIP-DKP})
    %         \end{minipage}%
    %     }
    % \end{minipage}

    \begin{minipage}[t][0.84cm][t]{0.42\textwidth} % 设置minipage的高度为4cm
        \colorbox{mycolor}{%
            \begin{minipage}{\dimexpr\textwidth-2\fboxsep} % 设置minipage的宽度
                ${\bm{\phi}}^{t+1}_{\textit{DKP}} = \bm{\phi}^{t}_{\textit{DKP}} + \frac{\delta^2}{2}  \frac{\partial \log p_{\textit{}}( \bm{\phi}^{t}_{\textit{DKP}}|{\bm{x}}^{t},\bm{y})}{\partial \bm{\phi}^{t}_{\textit{DKP}}}+
                \delta \frac{\partial \log p_{\textit{}}(\bm{\phi}^{t}_{\textit{DKP}}|{\bm{k}}^t_p)}{\partial \bm{\phi}^{t}_{\textit{DKP}}}$
            \end{minipage}%
        }
    \end{minipage}

    \vspace{-0.35em}
    
    \begin{minipage}[t][0.45cm][t]{0.42\textwidth} % 设置minipage的高度为4cm
        \colorbox{mycolor}{%
            \begin{minipage}{\dimexpr\textwidth-2\fboxsep} % 设置minipage的宽度
                ${\bm{k}}^{t+1} = \text{G}_{\textit{DKP}}({\bm{\phi}}^{t+1}_{\textit{DKP}})$
            \end{minipage}%
        }
    \end{minipage}

    \begin{minipage}[t][0.4cm][t]{0.42\textwidth} % 设置minipage的高度为4cm
        \colorbox{mycolor1}{%
            \begin{minipage}{\dimexpr\textwidth-2\fboxsep} % 设置minipage的宽度
                \% DIP-based image restoration stage
            \end{minipage}%
        }
    \end{minipage}
    
    \vspace{-0.08em}
    
    % \begin{minipage}[t][0.4cm][t]{0.42\textwidth} % 设置minipage的高度为4cm
    %     \colorbox{mycolor1}{%
    %         \begin{minipage}{\dimexpr\textwidth-2\fboxsep} % 设置minipage的宽度
    %             \State  $\log p_{\textit{}}( \bm{\phi}^{t-1}_{\textit{DIP}}|\bm{y},\bm{k}^{t}) = \|\bm{y}-(\text{G}_{\textit{DIP}}(\bm{\phi}^{t-1}_{\textit{DIP}})\otimes {\bm{k}}^{t})\downarrow _s\|_F^2,$
    %         \end{minipage}%
    %     }
    % \end{minipage}
    
    % \vspace{-0.08em}
    
    \begin{minipage}[t][0.8cm][t]{0.42\textwidth} % 设置minipage的高度为4cm
        \colorbox{mycolor1}{%
            \begin{minipage}{\dimexpr\textwidth-2\fboxsep} % 设置minipage的宽度
                ${\bm{\phi}}^{t+1}_{\bm{x}} = {\bm{\phi}}^{t}_{\bm{x}} + \gamma^{t}_{\bm{x}} \frac{\partial \log p_{\textit{}}( \bm{\phi}^{t}_{\bm{x}}|\bm{y},\bm{k}^t)}{\partial \bm{\phi}^{t}_{\bm{x}}}$
                % $\;$ (Eq. \eqref{eq:DIP-DKP RE loss})
            \end{minipage}%
        }
    \end{minipage}
    
    \vspace{-0.35em}
    
    \begin{minipage}[t][0.4cm][t]{0.42\textwidth} % 设置minipage的高度为4cm
        \colorbox{mycolor1}{%
            \begin{minipage}{\dimexpr\textwidth-2\fboxsep} % 设置minipage的宽度
                ${\bm{x}}^{t+1} = \text{G}_{\bm{x}}({\bm{\phi}}^{t+1}_{\bm{x}})$  
                % $\;$ (Eq. \eqref{eq:DIP-DKP-x}) 
            \end{minipage}%
        }
    \end{minipage}

    % \colorbox[RGB]{196,216,242}{\% DKP-based kernel estimation stage $\;\;\;\;\;\;\;\;\;\;\;\;\;\;\;\;\;$}

    % \colorbox[RGB]{196,216,242}{Sample random kernels $\{{\bm{k}}_r^{l}\}_{l=1}^L$ via MC. $\;\;\;\;\;\;\;\;\;\;\;\;$}

    % \colorbox[RGB]{196,216,242}{${\bm{k}}^t = {f}_{\text{DKP}}\left({\bm{x}}^{t-1},\bm{y},\{{\bm{k}}_r^{l}\}^L_{l=1}\right)$ $\;$ (Eq. \eqref{eq:DKP model in DIP-DKP}) $\;\;\;\;\;\;\;\;\;\;\;$}

    % \colorbox[RGB]{242,232,227}{\% DIP-based image restoration stage $\;\;\;\;\;\;\;\;\;\;\;\;\;\;\;\;\;\;\;$}

    % \colorbox[RGB]{242,232,227}{$\log p_{\textit{}}( \bm{\phi}^{t-1}_{\textit{DIP}}|\bm{y},\bm{k}^{t}) = \|\bm{y}-(\text{G}_{\textit{DIP}}(\bm{\phi}^{t-1}_{\textit{DIP}})\otimes {\bm{k}}^{t})\downarrow _s\|_F^2,$ }
    % % $\;$ (Eq. \eqref{eq:DIP network update})

    % \colorbox[RGB]{242,232,227}{${\bm{\phi}}^{t}_{\textit{DIP}} = {\bm{\phi}}^{t-1}_{\textit{DIP}} - \gamma^{t-1}_{\textit{DIP}} \cdot \frac{\partial \log p_{\textit{}}( \bm{\phi}^{t-1}_{\textit{DIP}}|\bm{y},\bm{k})}{\partial \bm{\phi}^{t-1}_{\textit{DIP}}}$ (Eq. \eqref{eq:DIP-DKP RE loss}) }

    % \colorbox[RGB]{242,232,227}{${\bm{x}}^{t} = \text{G}_{\textit{DIP}}({\bm{\phi}}^{t}_{\textit{DIP}})$  $\;$ (Eq. \eqref{eq:DIP-DKP-x}) $\;\;\;\;\;\;\;\;\;\;\;\;\;\;\;\;\;\;\;\;\;\;\;\;\;\;\;\;\;$}  % 206,100,103
    
    }
    \textbf{Output:} ${\bm{x}}^T$, ${\bm{k}}^T$.
\caption{\label{alg: DIP-DKP} The proposed DIP-DKP.}
\vspace{-0.03in}
\end{algorithm}

\noindent
\textbf{Proposed DIP-DKP.}
We replace the untrained or pre-trained networks for kernel priors in the existing DIP-based alternative framework by the proposed DKP model, which we refer to as DIP-DKP.
The objective of our proposed DIP-DKP can be formulated as follows,
% \begin{numcases}
% \;{\bm{k}}^t = {f}_{\text{DKP}}\left({\bm{x}}^{t-1},\bm{y}\right), \label{eq:DKP model in DIP-DKP}\\
% {\bm{\phi}}^{t}_{\textit{DIP}} = {\bm{\phi}}^{t-1}_{\textit{DIP}} + \gamma^{t-1}_{\textit{DIP}} \cdot \frac{\partial \log p_{\textit{}}( \bm{\phi}^{t-1}_{\textit{DIP}}|\bm{y},\bm{k}^t)}{\partial \bm{\phi}^{t-1}_{\textit{DIP}}},\label{eq:DIP-DKP RE loss}\\
% {\bm{x}}^{t} = \text{G}_{\textit{DIP}}({\bm{\phi}}^{t}_{\textit{DIP}}),\label{eq:DIP-DKP-x}
% \end{numcases}
\begin{numcases}
\;{\bm{\phi}}^*_{\bm{x}}, {\bm{\phi}}^*_{\textit{DKP}} = \underset{\bm{\phi}_{\bm{x}}, {\bm{\phi}}_{\textit{DKP}}}{\arg \min}\;\|\bm{y}- (\text{G}_{\textit{DKP}}({\bm{\phi}}_{\textit{DKP}}) \otimes  \notag\\
\;\;\;\;\;\;\;\;\; \text{G}_{\bm{x}}(\bm{\phi}_{\bm{x}}))\downarrow _s\|_F^2 + \|\text{G}_{\textit{DKP}}(\bm{\phi}_{\textit{DKP}}) - {\bm{k}}_p)\|_F^2,\label{eq:DIP-DKP loss} \\
{\bm{x}}^* = \text{G}_{\bm{x}}({\bm{\phi}}^*_{\bm{x}}), {\bm{k}}^* = \text{G}_{\textit{DKP}}({\bm{\phi}}^*_{\textit{DKP}}), \label{eq:DIP-DKP-x}
\end{numcases}
where $\text{G}_{\textit{DKP}}({\bm{\phi}}_{\textit{DKP}})$ is the kernel network of the proposed DKP model.
% where Eq. \eqref{eq:DKP model in DIP-DKP} denotes that the estimated kernel ${\bm{k}}^t$ is obtained by our DKP model, Eq. \eqref{eq:DIP-DKP RE loss} denotes that the DIP parameters are updated based on the LR image data-consistency.
% Eq. \eqref{eq:DIP-DKP-x} represents the estimated image ${\bm{x}}^t$ is obtained by the updated DIP network.

% \noindent
% \textbf{Framework.}
The overall solution procedure of the proposed DIP-DKP is given in Algorithm \ref{alg: DIP-DKP}.
% Let $t=1,2,\ldots,T$ be the index of the update iteration.
% At each instance of the BSR problem, 
% the LR image $\bm{y}$ is the given observation, and the DIP parameters $\bm{\phi}^0_{\textit{DIP}}$ are randomly initialized.
At each $t^{th}$ iteration, the kernel $\bm{k}^t$ is estimated in the DKP-based kernel estimation stage and then is assigned to the DIP model for HR image restoration in the forward propagation. 
In the back propagation, the parameters of DIP and DKP, i.e., $\bm{\phi}_{\bm{x}}$ and $\bm{\phi}_{\textit{DKP}}$, are updated while solving the BSR problem via an unsupervised inference.
With DKP, DIP-DKP realizes an adaptive kernel learning along the convergence trajectory of the BSR objective function, enabling accurate and dynamic kernel estimation. 
Therefore, without expensive labeled data and long training time in advance, DIP-DKP can estimate HR image and blur kernel simultaneously in a plug-and-play style. 
% A set of random kernels $\{{\bm{k}}^l_{r}\}_{l=1}^L$ are first sampled via the MC simulation.
% Based on the $\bm{x}^{t-1}$, $\bm{y}$ and $\{{\bm{k}}^l_{r}\}_{l=1}^L$, our DKP model can estimate the kernel $\bm{k}^t$ through our DKP model via \eqref{eq:DKP model in DIP-DKP}, which will be assigned to the next DIP-based image restoration stage.
% In the DIP-based image restoration stage, the DIP parameters $\bm{\phi}^t_{\textit{DIP}}$ can be updated via Eq. \eqref{eq:DIP-DKP RE loss}.
% The estimated image $\bm{x}^t$ can be obtained through the updated DIP network via Eq. \eqref{eq:DIP-x}, which will be assigned to the next iteration.
% After repeating the above process T times until the updates stop, 
% we can obtain the final estimated kernel ${\bm{k}}^T$ and image ${\bm{x}}^T$.
% In each iteration, firstly the $L$ hidden variables $\{\hat{\bm{\Sigma}}_{r}_l\}_l_{l=1}$ randomly sampled via the MC simulation correspond to $L$ random kernels.
% Since the \textcolor{blue}{importance} of each random kernel is calculated via Eq. \eqref{eq:task-specific kernel MC}, which re-weights each kernel,
% the kernel prior term $\zeta_{\bm{\phi}_{\bm{k}}}^t$ can be computed via Eq. \eqref{eq:network update via LD MCIS}.
% Then, the parameters $\bm{\phi}^t_{\bm{k}}$ are updated via Eq. \eqref{eq:network update via LD}, while the parameters $\bm{\phi}^t_{\textit{DIP}}$ are updated via Eq. \eqref{eq:DIP RE loss}.

\subsection{Diff-DKP}

\noindent
\textbf{Original DDPM Inference Process.}
Denoising diffusion probabilistic models (DDPM) \cite{ho2020denoising} defines a T-step forward process to add noise to data and a T-step reverse process to restore desired data from the noise. When an off-the-shelf DDPM $S_{\bm{\theta}}$ is applied to solve image restoration problem, the reverse process is implemented as inference process to estimate the high quality image as follows,
\begin{numcases}
\; \bm{x}_{0|t} = \frac{1}{\sqrt{\overline{\alpha}^t}}(\bm{x}_t-S_{\bm{\theta}}(\bm{x}_t,t)\sqrt{1-\overline{\alpha}^t}), \label{eq:diffusion step1}\\
\bm{x}_{t-1} \sim p(\bm{x}_{t-1}|\bm{x}_t,\bm{x}_{0|t}), \label{eq:diffusion step2}
\end{numcases}
where $\bm{x}_{0|t}$ denotes the estimated HR image $\bm{x}_0$ at the $t^{th}$ step, and $\overline{\alpha}^t$ is the hyper-parameter. To ensure that HR images $\bm{x}_0 \sim q(\bm{x})$ can be reconstructed from random noise $\bm{x}_T \sim \mathcal{N}(\bm{0}, \bm{I})$, the existing methods typically re-train \cite{saharia2022image} or fine-tune \cite{yi2023diff} the DDPM model  via supervised learning on LR-HR datasets, or provide ground truth kernel \cite{wang2022zero} to enroll task-specific knowledge for convergence guarantee. However, the performance of DDPM is unstable, even when trained by a large number of labeled dataset.  

\begin{algorithm}[t]
    \SetAlgoLined
    \footnotesize
    \textbf{Given:} $\bm{y}$, $\bm{\phi}^T_{_{\textit{DKP}}}$, $S_{\bm{\theta}}$ and ${\bm{x}}_T\sim \mathcal{N}(\bm{0},\bm{I})$.
    
    % \textbf{Initialize:} ${\bm{x}}^T\sim \mathcal{N}(\bm{0},\bm{I})$.
    
    \For{t $\gets$ T, T-1, $\ldots$, 1}{
    
    % \begin{minipage}[t][0.4cm][t]{0.42\textwidth} % 设置minipage的高度为4cm
    %     \colorbox{mycolor}{%
    %         \begin{minipage}{\dimexpr\textwidth-2\fboxsep} % 设置minipage的宽度
    %             \State  Sample random kernels $\{{\bm{k}}_r^{l}\}_{l=1}^L$ via MC.
    %         \end{minipage}%
    %     }
    % \end{minipage}
    
    % \begin{minipage}[t][0.5cm][t]{0.42\textwidth} % 设置minipage的高度为4cm
    %     \colorbox{mycolor}{%
    %         \begin{minipage}{\dimexpr\textwidth-2\fboxsep} % 设置minipage的宽度
    %             ${\bm{k}}^t = {f}_{\text{DKP}}\left({\bm{x}}^{t-1},\bm{y}\right)$ 
    %             % $\;$ (Eq. \eqref{eq:DKP module in Diff-DKP})
    %         \end{minipage}%
    %     }
    % \end{minipage}
    
    % \vspace{-0.08em}
    
    \begin{minipage}[t][0.4cm][t]{0.42\textwidth} % 设置minipage的高度为4cm
        \colorbox{mycolor1}{%
            \begin{minipage}{\dimexpr\textwidth-2\fboxsep} % 设置minipage的宽度
                \% Diffusion-based image restoration process
            \end{minipage}%
        }
    \end{minipage}
    
    \vspace{-0.08em}
    
    \begin{minipage}[t][0.6cm][t]{0.42\textwidth} % 设置minipage的高度为4cm
        \colorbox{mycolor1}{%
            \begin{minipage}{\dimexpr\textwidth-2\fboxsep} % 设置minipage的宽度
                $\bm{x}_{0|t} = \frac{1}{\sqrt{\overline{\alpha}^t}}({\bm{x}}_{t}-\mathcal{S}_{\bm{\theta}}(\bm{x}_t,t)\sqrt{1-\overline{\alpha}^t})$ 
                % $\;$ (Eq. \eqref{eq:diffusion step1})
            \end{minipage}%
        }
    \end{minipage}
    
    \vspace{-0.08em}
    
    \begin{minipage}[t][0.4cm][t]{0.42\textwidth} % 设置minipage的高度为4cm
        \colorbox{mycolor}{%
            \begin{minipage}{\dimexpr\textwidth-2\fboxsep} % 设置minipage的宽度
                \% DKP incorporated data consistency refinement
            \end{minipage}%
        }
    \end{minipage}
    
    \vspace{-0.08em}

    \begin{minipage}[t][0.78cm][t]{0.42\textwidth} % 设置minipage的高度为4cm
        \colorbox{mycolor}{%
            \begin{minipage}{\dimexpr\textwidth-2\fboxsep} % 设置minipage的宽度
                ${\bm{\phi}}^{t-1}_{_{\textit{DKP}}} = \bm{\phi}^{t}_{_{\textit{DKP}}} + \frac{\delta^2}{2}  \frac{\partial \log p_{\textit{}}( \bm{\phi}^{t}_{_{\textit{DKP}}}|{\bm{x}}_{0|t},\bm{y})}{\partial \bm{\phi}^{t}_{_{\textit{DKP}}}}+
                \delta \frac{\partial \log p_{\textit{}}(\bm{\phi}^{t}_{_{\textit{DKP}}}|{\bm{k}}^t_p)}{\partial \bm{\phi}^{t}_{_{\textit{DKP}}}}$
            \end{minipage}%
        }
    \end{minipage}

    \vspace{-0.08em}
    
    \begin{minipage}[t][0.45cm][t]{0.42\textwidth} % 设置minipage的高度为4cm
        \colorbox{mycolor}{%
            \begin{minipage}{\dimexpr\textwidth-2\fboxsep} % 设置minipage的宽度
                ${\bm{k}}^{t-1} = \text{G}_{\textit{DKP}}({\bm{\phi}}^{t-1}_{\textit{DKP}})$
            \end{minipage}%
        }
    \end{minipage}
    
    \begin{minipage}[t][0.7cm][t]{0.42\textwidth} % 设置minipage的高度为4cm
        \colorbox{mycolor1}{%
            \begin{minipage}{\dimexpr\textwidth-2\fboxsep} % 设置minipage的宽度
                $\hat{\bm{x}}_{0|t} = \bm{x}_{0|t} + \gamma^{t}_{\bm{x}} \frac{\partial \log p_{\textit{}}( \bm{x}_{0|t}|\bm{y},\bm{k}^{t-1})}{\partial \bm{x}_{0|t}}$ 
                % $\;$ (Eq. \eqref{eq:Diff-DKP-x}) 
            \end{minipage}%
        }
    \end{minipage}
    
    \vspace{-0.08em}
    
    \begin{minipage}[t][0.4cm][t]{0.42\textwidth} % 设置minipage的高度为4cm
        \colorbox{mycolor1}{%
            \begin{minipage}{\dimexpr\textwidth-2\fboxsep} % 设置minipage的宽度
                $\bm{x}_{t-1} \sim p(\bm{x}_{t-1}|\bm{x}_t,\hat{\bm{x}}_{0|t})$ 
                % $\;$ (Eq. \eqref{eq:diffusion step2})
            \end{minipage}%
        }
    \end{minipage}

    % \colorbox[RGB]{196,216,242}{\% DKP-based kernel estimation stage $\;\;\;\;\;\;\;\;\;\;\;\;\;\;\;\;\;$}

    % \colorbox[RGB]{196,216,242}{Sample random kernels $\{{\bm{k}}_r^{l}\}_{l=1}^L$ via MC. $\;\;\;\;\;\;\;\;\;\;\;\;$}

    % \colorbox[RGB]{196,216,242}{${\bm{k}}^{t} = {f}_{\text{DKP}}\left({\bm{x}}^{t},\bm{y},\{{\bm{k}}^l_{r}\}^L_{l=1}\right)$ $\;$ (Eq. \eqref{eq:DKP model in DIP-DKP}) $\;\;\;\;\;\;\;\;\;\;\;\;\;\;$}

    % \colorbox[RGB]{242,232,227}{\% Diffusion-based image restoration stage $\;\;\;\;\;\;\;\;\;\;$}

    % \colorbox[RGB]{242,232,227}{$\bm{x}^{0|t} = \frac{1}{\sqrt{\overline{\alpha}^t}}({\bm{x}}^{t}-\mathcal{S}_{\bm{\theta}}(\bm{x}^t,t)\sqrt{1-\overline{\alpha}^t})$  $\;$ (Eq. \eqref{eq:diffusion step1})  }

    % \colorbox[RGB]{242,232,227}{$\hat{\bm{x}}^{0|t} = \bm{x}^{0|t} - \gamma^{t}_{\bm{x}} \cdot \frac{\partial \log p_{\textit{}}( \bm{x}^{0|t}|\bm{y},\bm{k}^t)}{\partial \bm{x}^{0|t}}$   $\;$ (Eq. \eqref{eq:Diff-DKP-x}) $\;\;\;\;$}

    % \colorbox[RGB]{242,232,227}{$\bm{x}^{t-1} \sim p(\bm{x}^{t-1}|\bm{x}^t,\hat{\bm{x}}^{0|t})$  $\;$ (Eq. \eqref{eq:diffusion step2 form}) $\;\;\;\;\;\;\;\;\;\;\;\;\;\;\;\;\;\;$}

    }
    \textbf{Output:} ${\bm{x}}_0$, ${\bm{k}}^0$.
\caption{\label{alg: Diff-DKP} The proposed Diff-DKP.}
\vspace{-0.03in}
\end{algorithm}

% We propose to model an alternative update framework between the HR image $\bm{x}_{S_{\bm{\theta}}}$ and parameters $\bm{\phi}_{\textit{DKP}}$ based on the off-the-shelf diffusion model and our DKP model.
% The objective of our Diff-DKP can be formulated as follows,
% \begin{numcases}
% % \;\bm{x}^{0|t} = \frac{1}{\sqrt{\overline{\alpha}^t}}(\bm{x}^t-\mathcal{S}_{\bm{\theta}}(\bm{x}^t,t)\sqrt{1-\overline{\alpha}^t})\\
% \;{\bm{x}_0, {\bm{\phi}}^*_{\textit{DKP}} = \underset{\bm{x}_{0|t}, {\bm{\phi}}_{\textit{DKP}}}{\arg \min}\;\|\bm{y}- (\bm{x}_{0|t}}\otimes  \notag\\
% \;\;\;\;\text{G}_{\textit{DKP}}({\bm{\phi}}_{\textit{DKP}}))\downarrow _s\|_F^2 + \|\text{G}_{\textit{DKP}}(\bm{\phi}_{\textit{DKP}}) - \hat{\bm{k}}_{r})\|_F^2,\label{eq:Diff-DKP loss} \\
% {\bm{k}}^* = \text{G}_{\textit{DKP}}({\bm{\phi}}^*_{\textit{DKP}}), \label{eq:Diff-DKP-x-k}
% \end{numcases}
% where $\bm{x}_{0|t}$ is the estimated HR image by the diffusion model $S_{\bm{\theta}}$.
\noindent
\textbf{Proposed Diff-DKP.}
The instability of DDPM mainly comes from the training process that involves multiple image processing tasks.
In this case, the off-the-shelf diffusion model cannot concentrate on BSR objective, thus leading to image distortion and content mismatch.
To alleviate this issue, the proposed Diff-DKP incorporates the DKP model to provide task-specific data-consistency knowledge on the basis of the vanilla DDPM reverse iterations.   
% To achieve the update of $\bm{x}_{0|t}$ in Eq. \eqref{eq:Diff-DKP loss}, 
Specifically, an external DKP incorporated data consistency refinement of $\bm{x}_{0|t}$ is inserted between \eqref{eq:diffusion step1} and \eqref{eq:diffusion step2}, given by 
\begin{equation}\label{eq:Diff-DKP-x}
\; \hat{\bm{x}}_{0|t} = \bm{x}_{0|t} + \gamma^{t}_{\bm{x}} \frac{\partial \log p_{\textit{}}( \bm{x}_{0|t}|\bm{y},\bm{k}^t)}{\partial \bm{x}_{0|t}}, 
\end{equation}
where $\gamma^{t}_{\bm{x}}$ is the update step, and
\begin{equation}\label{eq:data-consis}
\; \log p_{\textit{}}( \bm{x}_{0|t}|\bm{y},\bm{k}^t) = - \|\bm{y}-(\bm{x}_{0|t}\otimes {\bm{k}}^{t})\downarrow _s\|_F^2, 
\end{equation}
which enables the inference process to converge to the right direction along with the data-consistent solution.

The overview of the Diff-DKP algorithm is presented in Algorithm \ref{alg: Diff-DKP}.
Let $t=T,T-1,\ldots,1$ denote the index of the diffusion reverse step. 
At each step, the diffusion model first estimates the $\bm{x}_{0|t}$.
Then, 
the DKP model adaptively generates kernel prior with respect to the latest $\bm{x}_{0|t}$, while $\bm{x}_{0|t}$ is further updated with respect to the data consistency Eq. \eqref{eq:data-consis}, thus, ensuring the inference process is underlying the BSR objective. 
It is noteworthy to point out that the parameters of the diffusion model are fixed and only the parameters of lightweight kernel estimator network are optimized in the inference process.  

In this way, the off-the-shelf diffusion model plays the role of HR image estimator, while the estimated HR image is further refined by the BSR task specific prior knowledge, referring to Eq. \eqref{eq:Diff-DKP-x}. Different from those methods that incorporate prior knowledge of BSR task via supervised re-training/fine-tuning, Diff-DKP behaves a plug-and-play scheme, thus without data demands and training cost before implementation. 

\section{Experiments}

\subsection{Experimental Setup} \label{sec:setup}

\begin{figure}[b]
  \centering
  \vspace{-0.5cm}
  \includegraphics[width=1\linewidth]{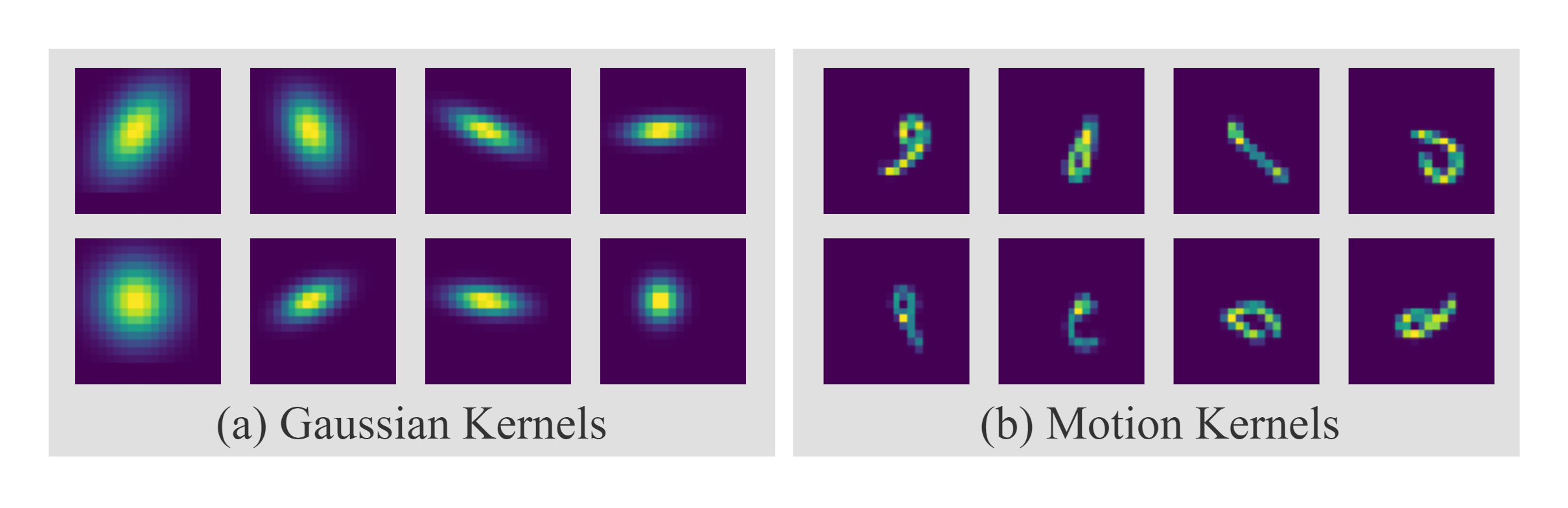}\\
  \vspace{-0.5cm}
  \caption{The visualization of the adopted blur kernels. }\label{fig:random kernel visual}
  \vspace{-0.3cm}
\end{figure}

\noindent
\textbf{Data Preparation.} 
Following the widely adopted kernel assumption \cite{riegler2015conditioned, wang2021unsupervised, liang2021flow, Yue2022blind}, we conduct  the experiments on anisotropic Gaussian kernels and motion kernels, which are shown in Fig. \ref{fig:random kernel visual}.
The kernel sizes are set to $(4s+3)\times(4s+3)$.
For the Gaussian kernel, the width ranges are set to $[0.175s, 2.5s]$, and the rotation angle range is set to $[0,\pi]$, with a scale factor $s = 4$, respectively.
For the motion kernel, we adopt random motion kernel generation method proposed by \cite{kupyn2018deblurgan}, which simulates realistic and complex blur kernels from random trajectories. Detailed formulations of Gaussian and motion kernels are given in the supplementary material. 
We synthesize LR images with random kernels with respect to Eq. (\ref{eq:degradation model}) for testing data based on five popular public benchmark datasets, including Set5 \cite{bevilacqua2012low}, Set14 \cite{zeyde2010single}, BSD100 \cite{martin2001database}, Urban100 \cite{huang2015single} and RealSRSet \cite{li2020efficient}.
We compare these kernels in terms of the peak signal to noise ratio (PSNR), and compare HR images in terms of PSNR and structural similarity (SSIM) \cite{wang2004image}.

\noindent
\textbf{Comparison Methods.}
The proposed DIP-DKP and Diff-DKP are compared with existing baselines including: Double-DIP \cite{ren2020neural}, DIP-FKP \cite{liang2021flow}, DASR \cite{wang2021unsupervised}, BSRDM \cite{Yue2022blind}, DCLS \cite{luo2022deep}, DARM \cite{zhou2023learning} and DiffBSR \cite{lin2023diffbir}.
Specifically, Double-DIP tends to provide kernel priors by training a FCN network only with respect to the LR image restoration error. 
DIP-FKP incorporates the FKP model as kernel prior which is pre-trained on kernel datasets. KernelGAN+ZSSR and DARM are self-supervised and train an interal generative adversarial network (GAN) to estimate the blur kernel. 
BSRDM formulates an elaborate degradation modelling on noise and kernel as handcrafted priors. 
DASR is a representative end-to-end method that is pre-trained on DIV2K \cite{agustsson2017ntire} and Flickr2K \cite{timofte2017ntire} HR image datasets. 
DiffBSR is fine-tuned on BSR labeled datasets before applied to estimate HR images.  
% Specifically, DIP, Double-DIP, KernelGAN+ZSSR, BSRDM and DARM are applied in a plug-and-play fashion without any training in advance as well as extra labeled data. DIP-FKP pre-trains their kernel estimation model on a kernel dataset, which fits the Gaussian kernel settings.
% DASR uses DIV2K \cite{agustsson2017ntire} and Flickr2K \cite{timofte2017ntire} as the HR image training set, while DiffBSR is trained on ImageNet \cite{deng2009imagenet}.

% \noindent
% \textbf{Networks.}
% In our DKP module as shown in Fig. \ref{fig: DKP} (c), the kernel estimation network $\text{G}_{\bm{k}}$ is a three-layer fully-connected network (FCN).
% The dimension of input layer and output layer is equal to the dimension of the blur kernel to be estimated, while the dimension of hidden layer is set to $500$.
% The input of the FCN is a random Gaussian noise, which is fixed in the test stage.
% In Fig. \ref{fig: DKP} (b), the image restorer of the SR module $\text{G}_{\textit{DIP}}$ is from the DIP model \cite{ulyanov2018deep}.
% It should be noted that the current image $\hat{\bm{x}}_\textit{old}$ is not necessary as an input for the DIP model. 

\noindent
\textbf{Implementation and Hyper-parameters.}
% There is no pre-training or training dataset for our DIP-DKP, in which parameters are randomly initialized and updated in the unsupervised way.
The adopted kernel estimation network $\text{G}_{\bm{k}}$ of PKE module in this paper is a three-layer fully-connected network (FCN).
% The dimension of input layer and output layer is equal to the dimension of the blur kernel to be estimated, while the dimension of hidden layer is set to $500$.
% The input of the FCN is a random Gaussian noise, which is fixed in the test stage.
% Note that there is no labeled dataset for our DIP-DKP and Diff-DKP due to that we do not need any pre-training for DIP-DKP and fine-tuning for DIP-DKP but directly use the untrained DIP model and pre-trained diffusion model. 
The adopted DIP model follows the original settings in \cite{ulyanov2018deep}, and the diffusion model is the vanilla version \cite{ho2020denoising} that is trained on ImageNet \cite{deng2009imagenet}.
% We choose the pre-trained model proposed by \cite{ho2020denoising}. 
% The network details are presented in Sec \ref{sec:SR Module}.
% Following the settings in \cite{ulyanov2018deep, wang2022zero}, $T$ in DIP-DKP is set to $1000$, $J$ in Diff-DKP is set to $100$.
The number of sampling times in the MCMC simulation $L$ is the only hyper-parameter in the proposed DKP model. 
The hyper-parameter tuning results are given in Table \ref{table:turning for L}.It is explicit that the performance reaches equilibrium around $L\in[4, 8]$. 
To balance the efficiency and effectiveness, we set $L=5$ in this paper.

\begin{table*}[t] 
\setlength{\abovecaptionskip}{0cm}
\setlength{\belowcaptionskip}{0cm}
\caption{Average image PSNR performance of the proposed DIP-DKP and Diff-DKP on Set5 \cite{bevilacqua2012low} on the Gaussian kernel scenario. 
% (Random Gaussian kernels with $s=4$)
}
\begin{center}\label{table:turning for L}
% \footnotesize
\scriptsize
\vspace{-0.1cm}
\begin{tabular}{ |m{2cm}<{\centering}|m{1cm}<{\centering} |m{1cm}<{\centering} |m{1cm}<{\centering} |m{1cm}<{\centering} |m{1cm}<{\centering} |m{1cm}<{\centering}|m{1cm}<{\centering}|}
\hline
Methods       &$L=0$     &$L=2$    &$L=4$     &$L=6$    &$L=8$   &$L=10$  &$L=15$ \\
\hline
\end{tabular}

\vspace{1.2pt}
\begin{tabular}{ |m{2cm}<{\centering}|m{1cm}<{\centering} |m{1cm}<{\centering} |m{1cm}<{\centering} |m{1cm}<{\centering} |m{1cm}<{\centering} |m{1cm}<{\centering}|m{1cm}<{\centering}|}
\hline
    {DIP-DKP} (Ours)     & 20.99   & 27.12   & 28.44  & \textbf{28.57}  & 28.52  & 28.29  & 28.03\\
% \hline
%     {Diff-DKP}     & $21.97$   & $31.95$   & $32.12$  & $32.19$  & $32.69$  & $32.50$ \\
% \hline
    {Diff-DKP} (Ours)    & 21.97   & 28.95   & 29.40  & 29.47  & \textbf{29.76}  & 29.67  & 29.26\\
\hline
\end{tabular}
\end{center} 
\vspace{-0.5cm}
\end{table*}

\begin{table*}[thbp] 
\setlength{\abovecaptionskip}{0cm}
\setlength{\belowcaptionskip}{0cm}
\caption{Average PSNR/SSIM of different methods on public datasets that are synthesized by the random Gaussian/Motion kernels with $s=4$. The best and second best results are highlighted in \textcolor{red}{red} and \textcolor{blue}{blue} colors, respectively.
% The gray results represent unfair comparisons due to mismatch of degradation models. The best results are emphasized with bold.
}
\vspace{-0.5cm}
\begin{center} \label{table:images PSNR Gaussian}
\scriptsize
\renewcommand{\arraystretch}{0.9}
\vspace{0.4cm}
\begin{tabular}{ |m{3.5cm}<{\centering} |m{1.2cm}<{\centering} |m{1.8cm}<{\centering} |m{1.8cm}<{\centering} |m{1.8cm}<{\centering} |m{1.8cm}<{\centering}|}
\hline
\text{Method}                    &  Kernel     &Set5 \cite{bevilacqua2012low}                   &Set14 \cite{zeyde2010single}                        &BSD100  \cite{martin2001database}                       &Urban100  \cite{huang2015single}\\
\hline
\end{tabular}

\vspace{1.2pt}
\begin{tabular}{ |m{3.5cm}<{\centering} |m{1.2cm}<{\centering} |m{1.8cm}<{\centering} |m{1.8cm}<{\centering} |m{1.8cm}<{\centering} |m{1.8cm}<{\centering}|}
\hline
    % Bicubic Interpolation        &Gaussian   & {$21.70/0.6198$}      & {$20.86/0.5181$}            & {$21.95/0.5097$}             & {$19.13/0.4729$}\\
    % DIP   \cite{ulyanov2018deep}        &Gaussian   & {$27.34/0.7465$}      & {$25.03/0.6371$}            & {$24.92/0.6030$}             & {$22.55/0.6128$}\\
    Double-DIP  \cite{ren2020neural}    &    & {20.99/0.5578}      & {18.31/0.4426}            & {18.57/0.3815}             & {18.15/0.4491}\\
    % KernelGAN\cite{gu2019blind}+ZSSR \cite{shocher2018zero}      &    & {24.46/0.6216}      & 22.65/0.5414           & 21.49/0.5229            & 21.04/0.4979\\
    DASR \cite{wang2021unsupervised}    &    & {27.37/0.7859}      & 25.43/0.6591           & 25.11/0.6129            & 22.88/0.6448\\
    DIP-FKP   \cite{liang2021flow}      &   & {27.77/0.7914}      & 25.65/0.6764            & {25.15/0.6354}             & {22.89}/0.6327\\
    BSRDM    \cite{Yue2022blind}        & Gaussian   & {27.81/0.8029}      & 25.35/{0.6859}            &{25.61/0.6526}             & 22.36/{0.6601}\\
    {DCLS}    \cite{luo2022deep}        & kernel     & {27.50/0.7948}   & {25.68/0.6639}   & {25.34/0.6169}   & {22.92/0.6475}\\
    DiffBIR \cite{lin2023diffbir}       & scenario          & {25.15/0.6468}      & 23.01/0.5935           &23.88/0.5586            & 21.94/0.5657\\
    DARM \cite{zhou2023learning}        &    & {26.25/0.6818}      & 24.19/0.6187           & 24.29/0.5898            & 22.14/0.5967\\
    {DIP-DKP} (Ours)             &   & \color{blue}{28.03/0.8039}      & \color{blue}{25.98/0.6878}           & \color{blue}{25.66/0.6531}             & $\color{blue}{23.24/0.6644}$\\
    {Diff-DKP} (Ours)            &   & \color{red}{29.44/0.8592}      & \color{red}{26.76/0.7400}            & \color{red}{26.63/0.7057}             & \color{red}{23.92/0.6875}\\
  \hline
\end{tabular}
% \end{center} 
% \vspace{-0.2cm}
% \end{table*}

% \begin{table*}[thbp] 
% \setlength{\abovecaptionskip}{0cm}
% \setlength{\belowcaptionskip}{0cm}
% \caption{Average PSNR/SSIM of different methods on various datasets that are synthesized by the random motion kernels. The best and second best results are highlighted in \textcolor{red}{red} and \textcolor{blue}{blue} colors, respectively.
% % The gray results represent unfair comparisons due to mismatch of degradation models. The best results are emphasized with bold.
% }
% \vspace{-0.2cm}
% \begin{center} \label{table:images PSNR motion}
% \footnotesize
% \renewcommand{\arraystretch}{1}
% % \setlength{\tabcolsep}{8pt}{
% \vspace{0.4cm}
% \begin{tabular}{ |m{4.3cm}<{\centering} |m{0.8cm}<{\centering} |m{2.1cm}<{\centering} |m{2.1cm}<{\centering} |m{2.1cm}<{\centering} |m{2.1cm}<{\centering}|}
% \hline
% \text{Method}                    &Scale         &Set5 \cite{bevilacqua2012low}                   &Set14 \cite{zeyde2010single}                        &BSD100  \cite{martin2001database}                       &Urban100  \cite{huang2015single}\\
% \hline
% \end{tabular}

\vspace{1.2pt}
\begin{tabular}{ |m{3.5cm}<{\centering} |m{1.2cm}<{\centering} |m{1.8cm}<{\centering} |m{1.8cm}<{\centering} |m{1.8cm}<{\centering} |m{1.8cm}<{\centering}|}
\hline
    % Bicubic Interpolation        &Motion   & {$21.42/0.6087$}      & {$21.26/0.5335$}            & {$20.98/0.5179$}             & {$19.47/0.4862$}\\
    % DIP   \cite{ulyanov2018deep}        &Motion   & {$23.35/0.6263$}      & {$23.05/0.5997$}            & {$21.95/0.5634$}             & {$20.45/0.5269$}\\
    Double-DIP  \cite{ren2020neural}    &   & {18.92/0.4510}      & {20.41/0.4847}            & {19.00/0.3757}             & {15.42/0.2932}\\
    % KernelGAN\cite{gu2019blind}+ZSSR \cite{shocher2018zero}      &   & {22.56/0.6141}      & {22.18/0.5573}            & {20.82/0.5239}             & {19.89/0.5023}\\
    DASR \cite{wang2021unsupervised}    &   & {24.21/0.7252}      & {24.16/0.6145}            & {22.47/0.5836}             & {20.24/0.5478}\\
    DIP-FKP   \cite{liang2021flow}      &    & {24.61/0.7371}      & 24.21/0.6227            & {22.80/0.5880}             & {20.33/0.5572}\\
    BSRDM    \cite{Yue2022blind}        & Motion   & {24.01/0.7098}      & 23.56/0.6009            & {22.62/0.5791}             & 20.40/0.5494\\
    DCLS      \cite{luo2022deep}        & kernel       &{24.78/0.7323}      & {24.38/0.6211}   & {22.74/0.5922}   & {20.49/0.5534}\\
    DiffBIR \cite{lin2023diffbir}       & scenario      & {23.63/0.6367}      & 23.59/0.6043            & {22.35/0.5784}             & {20.14/0.5347}\\
    DARM \cite{zhou2023learning}        &   & {24.23/0.7269}      & {23.95/0.6294}            & {22.48/0.5830}             & {20.58/0.5595}\\
    {DIP-DKP} (Ours)             &   & \color{blue}{25.30/0.7417}      & \color{blue}{24.52/0.6434}            & \color{blue}{23.02/0.6136}             & \color{blue}{21.24/0.5667}\\
    {Diff-DKP} (Ours)            &   & \color{red}{28.74/0.8313}      & \color{red}{26.03/0.6719}            & \color{red}{24.10/0.6287}            & \color{red}{22.26/0.5862}\\
  \hline
\end{tabular}
\end{center} 
\vspace{-0.6cm}
\end{table*}

\begin{figure*}[t]
\vspace{-0.3cm}
\setlength{\abovecaptionskip}{-0cm}
\setlength{\belowcaptionskip}{-0.3cm}
  \centering
  \includegraphics[width=0.9\linewidth]{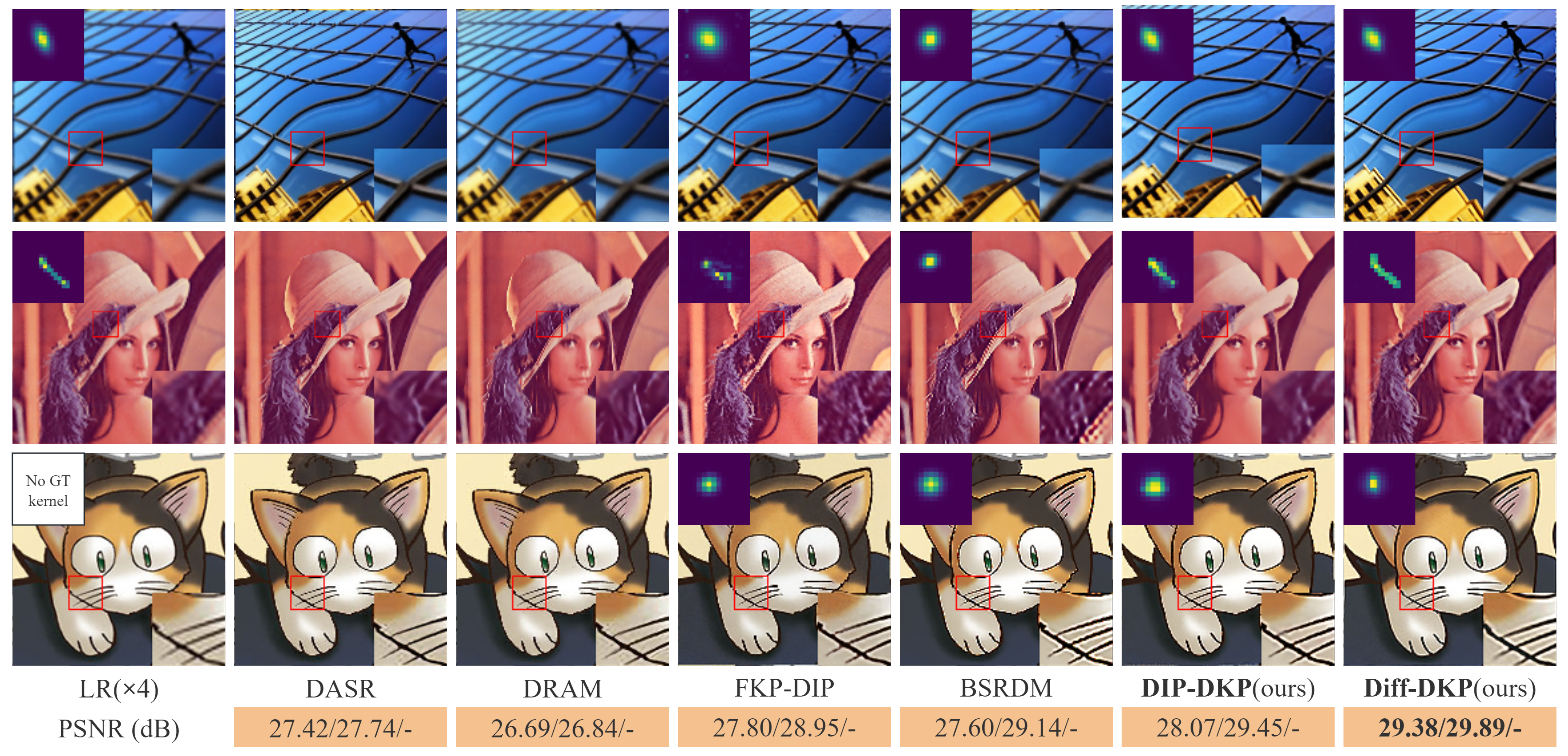}\\
  \vspace{-0cm}
  \caption{Visual results of different methods on public datasets for scale factor 4. Estimated/ground-truth kernels are shown on the top left.}\label{fig:visual results}
\end{figure*}

\begin{figure}[btph]
  \centering
  \vspace{-0.1cm}
  \includegraphics[width=1\linewidth]{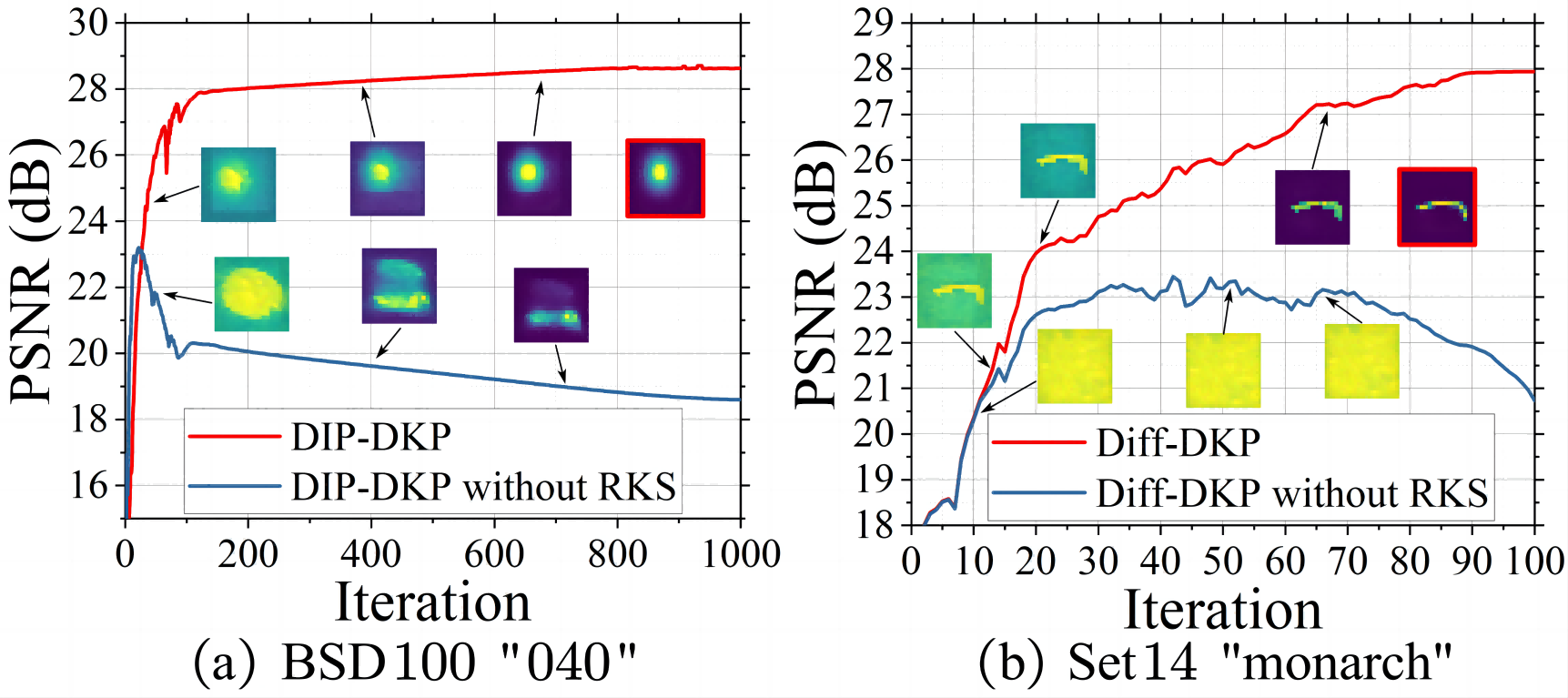}\\
  \vspace{-0.3cm}
  \caption{The intermediate results of DIP-DKP, Diff-DKP and their no RKS module versions over iterations on two test images.}\label{fig:ablation}
  \vspace{-0.2cm}\label{fig:ablations}
\end{figure}

\subsection{Comparison with State-of-the-Arts}

% We evaluate our DIP-DKP and Diff-DKP in the two main scenarios: the Gaussian kernels and motion kernels.

\noindent
\textbf{Evaluation on Gaussian Kernel Scenario.}
Quantitative evaluation results on four datasets with scale factors $s=4$ are presented in the upper half part of Table \ref{table:images PSNR Gaussian}.
We can see that the proposed DIP-DKP and Diff-DKP achieve the second and the best results on all datasets. 
We note that DIP-DKP only realizes slightly higher performance than the existing state-of-the-art (SotA) methods, while Diff-DKP achieves significantly better performances. 
This recalls our demonstrations in Section \ref{sec:SR Module}: DIP-DKP is totally trained while solving from scratch, and the DKP model plays the role of providing better convergence guarantee. 
Diff-DKP utilizes the DKP model to guide the well-trained diffusion model with fruitful data priors to converge to BSR task for better HR image restoration performances. 
In Table \ref{table: ablation}, we further show that our DKP model achieves the accurate kernel estimation with  higher kernel PSNR.
% Table \ref{table: ablation} shows the average kernel estimation performance of our DIP-DKP, Diff-DKP and other current state-of-the-art (SotA) kernel estimation methods.
% When comparing with the current state-of-the-art methods, DIP-FKP and BSRDM, the superior performance of kernel PSNR and image PSNR/SSIM of our DIP-DKP verifies that our DKP model exhibits better kernel estimation, since they all use the same DIP model to generate the HR image.
% Our Diff-DKP outperforms DIP-DKP significantly, suggesting that the diffusion-based image restoration model is more powerful and effective than the DIP-based image restoration model.

\noindent
\textbf{Evaluation on Motion Kernel Scenario.}
The lower half part of Table \ref{table:images PSNR Gaussian} shows the simulation results on the motion kernel scenario.
The supervised learning methods, i.e., DASR and DiffBIR, are re-trained/fine-tuned on motion kernel degraded HR image datasets. 
DIP-FKP is re-trained on the motion kernel dataset. 
The proposed DIP-DKP and Diff-DKP show significantly better performance on the motion kernel scenario, which validates that the proposed DKP model has good generalization ability towards different kernel categories. 
Specifically, Diff-DKP presents stable PSNR/SSIM scores when being applied to estimate motion kernels, while the rest suffer significant performance drop. This indicates that the proposed DKP is expected to handle kernel varying tasks.
% KernelGAN+ZSSR and DARM can adaptively learn the kernel priors through the adversarial loss.
% However, when the size of LR image is not large, they fail to achieve good image restoration results.
% DIP-FKP and BSRDM are designed to model the arbitrary Gaussian kernels.
% When the kernel changes from the Gaussian to motion, they show varying degrees of performance degradation.
% In contrast, our DIP-DKP and Diff-DKP only exhibit slight performance drop compared with other SotA methods in that case.
% This indicates that our DKP enjoys better generalization-ability, which are able to handle more complicated kernels due to its random kernel sampling modeling.

\noindent
\textbf{Visual Results.}
The visual results of different methods on synthetic and real-world images are shown in Fig. \ref{fig:visual results}.
We can see that 
1) In the case of Gaussian kernel, all methods are capable of producing satisfactory deblurring results,
while our DIP-DKP and Diff-DKP yield better results with more accurate kernel estimation.
2) In the case of motion kernel,
certain distortion on the estimated kernel can be seen in FKP-DKP and BSRDM fail to estimate motion kernel.
Meanwhile, our DIP-DKP and Diff-DKP achieve approximately accurate motion kernel estimation.
3) In the case of real image, both DIP-FKP and BSRDM estimate the Gaussian-like kernels, whereas our DIP-DKP and Diff-DKP tend to estimate the non-Gaussian kernels. 
This verifies that an adaptive and flexible kernel estimation discipline is learned by our DKP model, which may fit the real-world applications better.

\begin{table}[t] 
\setlength{\abovecaptionskip}{0cm}
\setlength{\belowcaptionskip}{0cm}
\caption{Average PSNR/SSIM of images and PSNR of kernels on Set14 \cite{zeyde2010single} with $s=4$. The best and second best results are highlighted in \textcolor{red}{red} and \textcolor{blue}{blue} colors, respectively.
% The gray results represent unfair comparisons due to mismatch of degradation models. The best results are emphasized with bold.
}
\vspace{-0.6cm}
\begin{center} \label{table: ablation}
\scriptsize
\renewcommand{\arraystretch}{0.9}
\vspace{0.4cm}
\begin{tabular}{ |m{3cm}<{\centering} |m{0.95cm}<{\centering} |m{0.7cm}<{\centering} |m{1.6cm}<{\centering}|}
\hline
\text{Method}                    &  Kernel      & Kernel PSNR                   & Image PSNR/SSIM \\
\hline
\end{tabular}

\vspace{1.8pt}
\begin{tabular}{ |m{3cm}<{\centering} |m{0.95cm}<{\centering} |m{0.7cm}<{\centering} |m{1.6cm}<{\centering}|}
\hline
    DIP-DKP without RKS      &     & 37.92      & 18.77/0.4227   \\
    Diff-DKP without RKS     &     & 40.93      & {17.33/0.3408}   \\
    Double-DIP \cite{ren2020neural}   &Gaussian     & {50.62}      & {18.31/0.4426}  \\
    DIP-FKP \cite{liang2021flow}      &kernel     & {54.46}      & {25.65/0.6764}  \\
    BSRDM \cite{Yue2022blind}        &scenario     & {55.38}      & 25.35/{0.6859}   \\
    {DIP-DKP} (Ours)             &     & \color{blue}{56.20}      &\color{blue}{25.98/0.6878}  \\
    {Diff-DKP} (Ours)            &     & \color{red}{56.76}      & \color{red}{26.76/0.7400}    \\
  \hline
\end{tabular}

\vspace{1.8pt}
\begin{tabular}{ |m{3cm}<{\centering} |m{0.95cm}<{\centering} |m{0.7cm}<{\centering} |m{1.6cm}<{\centering}|}
\hline
    DIP-DKP without RKS         &     & 34.92      & {18.19/0.4223}   \\
    Diff-DKP without RKS        &     & 34.78      & {17.65/0.3513}    \\
    Double-DIP \cite{ren2020neural}   &Motion     &35.52      & {20.41/0.4847}   \\
    DIP-FKP \cite{liang2021flow}      &kernel     & 37.52      & {24.21/0.6227}   \\
    BSRDM \cite{Yue2022blind}        &scenario     & 37.88      & {23.56/0.6009}   \\
    {DIP-DKP} (Ours)             &    & \color{blue}{39.33}      & \color{blue}{24.52/0.6434}   \\
    {Diff-DKP} (Ours)            &    & \color{red}{40.37}     & \color{red}{26.03/0.6719}     \\
  \hline
\end{tabular}
\end{center} 
\vspace{-0.6cm}
\end{table}

\subsection{Ablation Studies}
\noindent
\textbf{Ablation study of RKS module}. The ablation studies are carried on the MCMC sampling of kernel priors.
``Without RKS" denotes that the adopted DKP updates the kernel network only by the data-consistency term without the learned kernel prior.
In Fig. \ref{fig:ablations} (left), it can be seen that the estimated kernels without RKS have significant distortion, leading to remarkable PSNR drop of the estimated HR image, while DIP-DKP can estimate Gaussian kernels precisely with respect to the ground truth (with red frame).
Fig. \ref{fig:ablations} (right) shows that the accurate motion kernel estimation no longer exists when the RKS module is absent. It is thus obvious that without the kernel prior learned from the MCMC process, the Diff-DKP fails to converge to a rational motion kernel estimation.
The average kernel and image results are shown in Table \ref{table: ablation}.
Without kernel prior learned from the RKS module, the kernel estimation performances of DKP-based BSR methods have a significant performance drop, leading to poor image restoration performance as well.

\noindent
\textbf{Ablation study of PKE module}.
Since PKE essentially estimates blur kernels on the basis of the random kernel priors and LR observations, thus it is indispensable and we conduct ablation study on the different structures of kernel network in PKE module in Table \ref{table: PKE ablation study}.
We find that the kernel network performs well when it has 3-7 layers with 100-1000 units in each layer.
This indicates that the kernel network has good generalization-ability for the structure without the necessity of elaborately designing the network.

\begin{table}[pt!] 
\setlength{\abovecaptionskip}{0cm}
\setlength{\belowcaptionskip}{0cm}
\vspace{-0cm}
\caption{{The ablation of PKE module. (Set5, x4, image PSNR)}}
\begin{center} 
\scriptsize
\renewcommand{\arraystretch}{0.9}
% \tiny
\vspace{-0.2cm}
\begin{tabular}{ m{1.5cm}<{\centering}| m{1.1cm}<{\centering}| m{1.1cm}<{\centering}| m{1.1cm}<{\centering}|m{1.1cm}<{\centering}}
\hline
Layers$\backslash$Units     & 10     & 100     & 1000      & 10000         \\
\hline
1      & 13.75     & 23.57      & 28.93     & 28.24         \\

3      & 13.61     & \textbf{28.97}      & 28.48     & 28.35         \\

5      & 13.30     & 28.81      & 28.52     & 26.65         \\

7      & 13.86     & 28.30      & 28.54     & 27.93 \\
\hline
\end{tabular}
\label{table: PKE ablation study}
\end{center} 
\vspace{-0.8cm}
\end{table}

\subsection{Model Size, Runtime and Memory Usage}
The kernel network of our DKP model has a total of $562K$ parameters (FLOPs: $536K$) while Double-DIP and DIP-FKP have $641K$ parameters (FLOPs: $600K$) and $143K$ parameters (FLOPs: $178K$), respectively.
The runtime and memory usage of our DIP-DKP on a GeForce RTX 3090 GPU for generating a HR image of size $512\times512$ are about $92$ seconds and $11$GB memory, which is comparable with the Double-DIP ($91$ seconds and $11.2$GB) and DIP-FKP ($90$ seconds and $10.6$GB).
As for Diff-DKP, the $512\times512$ image needs to be divided into four $256\times256$ images for restoration, which costs a total of $60$ seconds and $4$GB memory.
Considering that our DIP-DKP and Diff-DKP are unsupervised and plug-and-play, it is reasonable to say that our methods have moderate computational costs.

Due to the page limitations, more experimental results are given in the supplementary material.
\section{Conclusion}
In this paper, we propose a dynamic kernel prior (DKP) model to solve the BSR problem in an unsupervised and pre-training-free paradigm. 
DKP realizes the rational kernel prior learning from MCMC sampling on random kernel distributions, providing accurate kernel estimation and thus leading to better HR image restoration. 
DKP can be easily incorporated with existing image restoration model, such as DIP and diffusion model, by replacing their kernel modeling modules or adding as an external kernel prior generator. 
When applied to solve the BSR problem, DKP is trained while solving the task with respect to the LR image restoration error, enabling no training necessity and labeled data demands. 
Extensive experiments on Gaussian and motion kernel scenarios with synthetic LR images and real-world images validate that DKP-based methods improve the kernel estimation accuracy significantly and thus lead to superior BSR results.
% we have proposed a novel kernel estimation model DKP, which consists of two main modules: the RKS module and PKE module.
% In RKS module, the kernel prior can be modeled by randomly sampled kernels based on the MCMC modeling.
% In PKE module, the rational and data-consistent kernel solution is guaranteed by the proposed network-based Langevin dynamics update strategy, which combines kernel prior from RKS module and LR image data-consistency.
% Then, we incorporate our kernel estimation model DKP into off-the-shelf image restoration models as two BSR methods: DIP-DKP and Diff-DKP, which achieve state-of-the-art performance.
% Most strikingly, both of them are unsupervised without any requirements on training in advance or labeled datasets.
We believe that the concept of using a trainable sampling process to provide adaptive priors will lead to a new direction in solving low-level tasks, aiming to achieve superior performance with modest computational costs in the way of unsupervised inference.
\vspace{-1cm}
{\small\bibliographystyle{ieeenat_fullname}
\bibliography{main}
}

% WARNING: do not forget to delete the supplementary pages from your submission 
% \input{sec/X_suppl}

\end{document}